\documentclass[prd,aps,showpacs,floats,letterpaper,floatfix,groupedaddress,eqsecnum,nofootinbib]{revtex4}

\usepackage{amssymb,amsmath}
\usepackage{graphicx,epsfig}
\usepackage{bm}

\newcommand{\be}{\begin{equation}}
\newcommand{\ee}{\end{equation}}
\newcommand{\bea}{\begin{eqnarray}}
\newcommand{\eea}{\end{eqnarray}}
\newcommand{\ba}{\begin{array}}
\newcommand{\ea}{\end{array}}
\newcommand{\beqn}{\begin{eqnarray*}}
\newcommand{\eeqn}{\end{eqnarray*}}

\def\nn{\nonumber}
\def\pa{\partial}
\def\ii{{{\rm i}\,}}

\def\nn{\nonumber}

\def\ii{{\rm i}}

\newlength{\sizeonefig}
\newlength{\sizetwofig}
\begin{document}

\title{Transformation of the multipolar components of gravitational radiation
  \\ under rotations and boosts}
\author{ Leonardo Gualtieri$^{1} \footnote{Electronic address:
    Leonardo.Gualtieri@roma1.infn.it}$,
Emanuele Berti$^{2} \footnote{Electronic address:
  Emanuele.Berti@jpl.nasa.gov}$,
Vitor Cardoso$^{3,4} \footnote{Electronic address:
  Vitor.Cardoso@ist.utl.pt}$
Ulrich Sperhake$^{5} \footnote{Electronic address:
  Ulrich.Sperhake@uni-jena.de}$ }
\affiliation{${^1}$ Dipartimento di Fisica, Universit\`a di Roma
``Sapienza'' and Sezione INFN Roma1, P.A. Moro 5, 00185, Roma, Italy}
\affiliation{${^2}$ Jet Propulsion Laboratory, California Institute of
  Technology, Pasadena, CA 91109, USA}
\affiliation{${^3}$ Centro Multidisciplinar de Astrof\'{\i}sica - CENTRA,
  Dept. de F\'{\i}sica, Instituto Superior T\'ecnico, Av.  Rovisco Pais 1,
  1049-001 Lisboa, Portugal}
\affiliation{${^4}$ Department of Physics and Astronomy, The University of
  Mississippi, University, MS 38677-1848, USA}
\affiliation{${^5}$ Theoretisch Physikalisches Institut, Friedrich Schiller
  Universit\"at, 07743 Jena, Germany}
\begin{abstract}
  We study the transformation of multipolar decompositions of gravitational
  radiation under rotations and boosts. Rotations to the remnant black hole's
  frame simplify the waveforms from the merger of generic spinning black hole
  binaries. Boosts may be important to get an accurate gravitational-wave
  phasing, especially for configurations leading to large recoil velocities of
  the remnant.  As a test of the formalism we revisit the classic problem of
  point particles falling into a Schwarzschild black hole. Then we highlight
  by specific examples the importance of choosing the right frame in numerical
  simulations of unequal-mass, spinning binary black-hole mergers.
\end{abstract}



\maketitle


\section{Introduction}

Multipolar expansions are a fundamental tool in the study of gravitational
radiation.  Such expansions can be performed using different sets of basis
functions (see \cite{Thorne:1980ru} for a review). It is common practice in
numerical relativity simulations to decompose the Weyl scalar $\Psi_4$ (a
quantity related to outgoing gravitational radiation) as a sum over the
Newman-Penrose {\em spin-weighted spherical harmonics} (SWSHs) of spin-weight
$s$, commonly denoted by $_sY_{lm}$. SWSHs reduce to ordinary spherical
harmonics for $s=0$, and $|s|=2$ is the spin-weight of interest for
gravitational radiation \cite{Newman:1966ub,Goldberg:1966uu}.

Compact binaries are one of the most promising candidate sources of detectable
gravitational waves for Earth-based and space-based interferometers.  The
post-Newtonian approximation, an analytical expansion of Einstein's equations
valid for orbital velocities $V/c\ll 1$, is usually employed to describe the
early inspiral of compact binaries. The post-Newtonian expansion becomes
increasingly inaccurate late in the inspiral, when orbital velocities become
large. At these late stages, time evolutions of compact binaries in numerical
relativity are necessary. High-accuracy numerical evolutions are difficult for
various reasons, including the approximate nature of the initial data, errors
arising from discretization and boundary effects.  Post-Newtonian expansions
and numerical calculations must be cross-validated by systematic comparisons
of the resulting waveforms, in order to determine a common region of validity
and to build reliable phenomenological templates
\cite{Baker:2006ha,Pan:2007nw,Buonanno:2007pf,Boyle:2007ft,Hannam:2007ik,Damour:2007yf,Ajith:2007qp,Ajith:2007kx,Yunes:2008tw}.
Since numerical waveforms are usually decomposed in SWSHs, significant efforts
have recently been devoted to study multipolar SWSH expansions both within
post-Newtonian theory \cite{Kidder:2007rt,Blanchet:2008je} and in connection
with numerical simulations
\cite{Buonanno:2006ui,Berti:2007fi,Schnittman:2007ij,Vaishnav:2007nm,Berti:2007nw}.

This paper deals with the transformation properties of SWSH decompositions of
gravitational radiation under rotations and boosts.  We will mostly focus on
the radiation emitted during the inspiral, merger and ringdown of compact
binaries, but our main results are general enough that they should be useful
in more general situations.

The transformation properties of gravitational waves have been studied as part
of the general problem of spacetime transformations in the sixties
\cite{Bondi:1962px,Sachs:1962zza,Sachs:1962wk,Newman:1966ub}. It was realized
that the asymptotic symmetries of asymptotically flat spacetimes form an
infinite-dimensional group, now known as the Bondi-Metzner-Sachs (BMS) group,
and that Lorentz transformations are a subgroup of the BMS group. Later on,
the transformations of SWSHs under the BMS group have been investigated in a
classic paper by Goldberg and collaborators \cite{Goldberg:1966uu}.

Unfortunately these pioneering works do not give an explicit prescription to
transform gravitational waves under rotations and boosts, and to handle their
multipolar decomposition. Our main purpose is to fill this gap and to provide
examples of the formalism at work, by considering specific problems in
numerical relativity and black-hole perturbation theory.
We only consider the emission as seen by an observer far away from the source,
and we assume the spacetime to be asymptotically flat.  As discussed in the
next section, the spacetime can then be described by a set of coordinates
$(u,r,\theta,\phi)$, where $\theta$ and $\phi$ are coordinates on the sphere,
$u$ is a retarded time parameter and the limit $r\rightarrow\infty$ (with $u$
finite) represents future null infinity. In this limit $r$ is much larger than
$u$ and any translation $\vec{\delta x}$ of the source:
$|\vec{\delta x}|/r\ll 1$, $|u|/r\ll 1$.
Furthermore, when we consider boosts we will always neglect quantities of
order $(V/c)^2$, where $V$ is the boost velocity of the source. This
approximation is well justified: according to recent results from numerical
relativity, black hole mergers produce recoil velocities $V\lesssim
4000$~km$/$s, i.e. $V/c\lesssim 0.013$
\cite{Campanelli:2007ew,Campanelli:2007cga,Gonzalez:2007hi,Pollney:2007ss,Dain:2008ck}.

In the remainder of this introduction we list some of the main motivations for
our work, and provide a brief summary of our most useful results.

\subsection{Motivations}

The multipolar structure of gravitational radiation is an important element in
the problem of detection and parameter estimation from compact binary
inspirals. The energy distribution determines which $(l,m)$ components of the
radiation are most easily detectable.  For a binary inspiral, it has long been
believed that the $l=m=2$ mode should dominate the energy emission
\cite{Flanagan:1997sx}. Numerical simulations confirm that the $l=m=2$ mode
carries $\sim 98\%$ of the radiation for equal-mass, non-spinning binaries
\cite{Buonanno:2006ui}.  However, higher modes can carry more than $\sim 10\%$
of the energy for unequal-mass binaries \cite{Berti:2007fi} and/or when the
black holes are spinning \cite{Vaishnav:2007nm,Berti:2007nw}.  Now, recall
that the gravitational-wave energy scales with the square of the waves'
amplitude.  Since SWSHs are normalized over the sphere, this means that the
amplitude ratio of the $l=m=3$ and $l=m=2$ modes is {\em typically} $\approx
1/3$ for generic binaries when different multipolar components are averaged
over the sky.  Depending on the relative direction of the source and the
detector, the enhancement of multipoles with $l>2$ can be even larger (see
\cite{Berti:2007zu} for a discussion in the context of ringdown detection).
For generic binary inspirals, a good phase accuracy on the $l=m=2$ mode is
necessary but not sufficient to produce accurate templates: higher multipolar
contributions {\em must} be taken into account to carefully track the phasing
of the waves over many inspiral cycles. One may expect that fewer multipolar
components should be sufficient in the early inspiral phase, but this
expectation is not confirmed by calculations. For example, by comparing
multipolar expansions of the (numerically computed) adiabatic energy flux to
post-Newtonian expansions of the flux in the context of extreme mass-ratio
inspirals, Ref.~\cite{Yunes:2008tw} showed that the number of multipoles
required to achieve a given post-Newtonian accuracy actually {\it increases}
in the Newtonian regime (i.e. for low orbital frequencies).

The transformation of multipolar decompositions of gravitational radiation
under rotations deserves a detailed study for many reasons, some of which we
list below.

{\em 1) Provide translation rules between numerical codes using different
  conventions.} A pressing motivation for our work is of practical nature.  As
more accurate and complex numerical waveforms are produced, it is vital to
establish a set of conventions between different numerical groups, and rules
to translate between different conventions. Binary simulations often differ in
subtle details of the numerical settings: for example, some groups initially
place the black holes on the $y$-axis, while other groups place them on the
$x$-axis.  This is obviously a matter of convention, but conventions are
important when setting up data formats to be used in LIGO and Virgo data
analysis (see e.g.~\cite{Brown:2007jx} for an attempt in this
direction). Similar issues should also be taken into account in collaborative
efforts to inject numerical waveforms in the LIGO data stream, such as the
\url{Ninja} project \cite{ninja}. One purpose of this paper is to provide a
``dictionary'' to translate between different conventions by relating
multipolar expansions in different frames.

{\em 2) Identify the optimal frame to study the merger/ringdown phase in
  generic spinning binary simulations.}  Given a generic configuration for a
spinning, inspiralling binary, at present there is no reliable analytical
method to predict (given the initial data) the orientation of the remnant's
spin. We will refer to the frame where the initial orbital angular momentum
points in the $z$-direction as the {\em simulation frame}, and to the frame
where the remnant black hole's spin points in the $z$-direction as the {\em
  remnant frame}.

In the simple (and most-studied) case of non-spinning binaries, a distinction
between the simulation frame and the remnant frame is not necessary.  The
black holes usually orbit on the $x-y$ plane, the initial orbital angular
momentum is in the $z$-direction, and angular momentum is radiated only in the
$z$-direction. Then the spin of the final black hole will also point in the
$z$-direction, and its dimensionless magnitude $j=J/M^2$ can be computed by
evaluating the $z$-component of the radiated angular momentum and using
balance arguments.
Since the remnant's spin points in the $z$-direction, a fit of the ringdown
frequencies for the dominant SWSH component of the waves {\em in the
  simulation frame} can then be used to verify the accuracy of wave extraction
methods \cite{Buonanno:2006ui,Berti:2007fi}.

For generic inspirals, however, the system will radiate angular momentum in
the $x$- and $y$-directions as well. The spin of the final black hole will
{\em not} be aligned with the $z$-axis of the simulation frame. 
If we (incorrectly) perform the SWSH decomposition in the simulation frame,
rather than in the remnant frame, the expansion will mix quasinormal modes
with the given value of $l$ and different values of $m$. For this reason,
using the wrong frame produces errors in the quasinormal frequencies of the
final black hole and in ringdown-based estimates of the value of the final
angular momentum.
Note that modes with {\em different values of $l$} are decoupled, since SWSHs
with fixed $l$ and $s$ form an irreducible representation of the rotation
group. In Section \ref{numerical} we will illustrate this problem by
specific examples, and we will show that rotations can be used to simplify the
waveforms: as viewed from the normal to the final black hole's rotation plane,
the merger-ringdown radiation {\em in the remnant frame} is circularly
polarized.

{\em 3) Measure the spin direction of the remnant through merger/ringdown
  waves.} The measurement of a single complex quasinormal frequency with
Earth-based \cite{Berti:2007zu} and space-based interferometers
\cite{Berti:2005ys} allows a determination of the mass and spin magnitude of
the hole resulting from a merger (the $l=m=2$ fundamental quasinormal mode is
again expected to be dominant). For Kerr black holes all quasinormal
frequencies are uniquely determined by the mass and spin of the hole, and the
measurement of the next subdominant mode (typically the fundamental mode with
$l=m=3$) will provide a test of the Kerr nature of the merger remnant.  Since
different multipolar components have different angular dependence, a
multi-mode detection could even be used to measure the {\em spin direction} of
the remnant. A detailed analysis of this problem requires the knowledge of the
transformation properties of the multipoles under rotations and boosts, and it
is an interesting topic for future work.

As argued earlier, boost velocities in binary mergers are typically small
($V/c\lesssim 0.013$). However the study of boosts is of more than academic
interest. In numerical simulations, gravitational radiation should ideally be
extracted on a sphere at null infinity. In practice the extraction sphere is
usually located at sufficiently large, but finite, distance from the source.
In this paper we point out that the distortion of the radiation pattern
induced by a large kick of the remnant can induce a mode mixing in the signal.
To leading order the amplitude of this mixing is proportional to the kick
velocity. The mixing effect will typically be small ($V/c \sim 10^{-2}$ for
the largest predicted kick velocities), but it may still affect the phasing
error of the inspiral waves over long inspiral times. Conversely, the
magnitude of the boost-induced mode mixing may set a threshold on the number
of high-order post-Newtonian terms we can (or should) compute to get an
accurate phasing. Finally, one should not exclude {\em a priori} the
possibility that the large recession velocities due to Hubble flow may produce
observable effects on gravitational waveforms.

Boosts are particularly relevant for the extraction of gravitational waves
from the collision of highly relativistic black holes with relatively small
impact parameter. This process is of interest as an extreme testing ground of
strong-field gravity and in the context of mini black hole production at the
Large Hadron Collider, and it is currently under investigation
\cite{Sperhake:2008ga}.

\subsection{Main results}

For the sake of clarity, and in view of possible generalizations of our
results, in the body of this paper we give a detailed account of our
calculations.  To help the reader, here we provide a short list of results
that we consider most useful in phenomenological applications:

\begin{itemize}

\item
  Eq.~(\ref{BMSpsi4transf0}) relates the Lorentz transformation of spin-$2$
  SWSHs to the transformation of the Weyl scalar $\Psi_4$.

\item
  Eq.~(\ref{sterharmall}) can be used to obtain spin-$2$ SWSHs in
  stereographic coordinates for general values of $l$, and
  Eq.~(\ref{sterharm}) gives closed-form expressions valid for $l=2$ and
  $l=3$.

%

\item
  For pure rotations different $l's$ do not couple because SWSHs with given
  $(s\,,l)$ form an irreducible representation of the rotation group, and we
  can consider the transformation laws {\em separately} for each value of $l$.
  The transformation of the harmonics under general rotations, in terms of a
  constant {\it rotation matrix}, is given in Eq.~(\ref{rotl}).

\item
  Eq.~(\ref{defAlmpm}) is one of our main results: it provides an expression
  for the general rotation matrix, valid for all values of $l$.
  Eqs.~(\ref{Ammprot}) and (\ref{l3general}) list components of the rotation
  matrix for $l=2$ and $l=3$, respectively.  Eqs.~(\ref{transforml2}) and
  (\ref{l3pi2}) further specialize these results to frames related by a
  $\pi/2$ rotation (i.e. $\theta\to \theta'=\theta-\pi/2$)
  The transformation under rotations of the Weyl scalar $\Psi_4$ and of its
  complex conjugate has a particularly simple form, which is given in
  Eqs.~(\ref{Cprime}), (\ref{Cprimeminus2}).

\item
  Eq.~(\ref{boost}) provides the $l=2$ transformation matrix under
  non-relativistic boosts, and Eq.~(\ref{boostmix}) shows in a special case
  how boosts produce mode-mixing.  As far as we know, this fact has never been
  pointed out in the existing literature.

\item
  In Section \ref{headon} we show that our transformation matrices are
  consistent with two ``classic'' calculations. In the first calculation we
  consider the equal-mass head-on collision of black holes as seen in two
  frames related by a $\pi/2$ rotation (i.e. $\theta\to
  \theta'=\theta-\pi/2$). Then we look at particles falling radially into
  Schwarzschild black holes along the $z$-axis or along radial, equatorial
  geodesics. We show that the SWSH components of the stress-energy tensor in
  these two frames are related by proportionality constants which are
  consistent with our transformation laws.

\item
  In Section \ref{genericrot} we demonstrate that our work can be useful to
  study the generic merger of unequal-mass, spinning black holes. We suggest a
  simple recipe to switch from the simulation frame to the remnant frame. Then
  we show that the merger waveform does indeed look simpler in the remnant
  frame, because in this frame: (1) ringdown radiation is circularly
  polarized, and (2) ringdown frequencies are consistent with the final black
  hole spin predicted from balance arguments. These properties do {\em not}
  apply to the simulation frame.

\end{itemize}

The paper is organized as follows. In Section \ref{BMSg} we introduce some
basic notions on the BMS group. Section \ref{BMStrans} is the core of this
paper, where we derive our main results on the transformation properties of
SWSHs and of gravitational radiation under rotations and boosts. Section
\ref{numerical} applies the formalism to specific problems in numerical
relativity and perturbation theory, and Section \ref{conclusions} discusses
some directions for future research. The Appendices contain various technical
topics: the expression of non-relativistic boosts in the context of BMS
transformations (Appendix \ref{boostapp}); the description of rotations and
boosts in the stereographic representation (Appendix \ref{stereoapp}); a
discussion of our numerical method to compute the radiated angular momentum
for generic, spinning black-hole binaries (Appendix \ref{radiatedj}); and a
brief summary of perturbation theory calculations of particles falling into
black holes (Appendix \ref{particleapp}).  Unless mentioned otherwise,
throughout the paper we use geometrical units ($G=c=1$).

\section{The Bondi-Metzner-Sachs group}
\label{BMSg}

To begin with, in this section we briefly recall the main steps of the
BMS-group construction. We refer to
\cite{Newman:1966ub,Sachs:1962zza,mccarthy:1837,Wald:1984cw,Carmeli:2000af}
for a more detailed and rigorous account.
Any asymptotically flat spacetime can be described by standard coordinates
$(u,r,\theta,\phi)$ defined as follows \cite{Sachs:1962zza}:

\begin{itemize}
\item $u$ is a retarded time parameter, defined by requiring that the
  hypersurfaces $u={\rm constant}$ are null: $g^{\mu\nu}u_{,\mu}u_{,\nu}=0$.
  These hypersurfaces are the wavefronts of gravitational waves emitted by the
  source.
\item $r$ is an affine parameter of the null geodesics tangent to
  $l^\mu=g^{\mu\nu}u_{,\nu}$ ($l^\mu$ is also a vector of the Newman-Penrose
  tetrad).
\item $\theta$ and $\phi$ are polar coordinates on the manifolds
  (diffeomorphic to the $2$-sphere) $r={\rm constant}$, $u={\rm constant}$.
  Geodesics with constant $(u,\theta,\phi)$ are called ``rays,'' and they are
  a congruence of geodesics, generators of the hypersurfaces $u={\rm
    constant}$.  The limit $r\rightarrow\infty$ with finite $u$ corresponds to
  future null infinity, where the gravitational radiation emitted by the
  source is detected.
\item Far away from the source one can define a time coordinate $t=u+r$.
\end{itemize}

The spacetime is asymptotically flat if the metric tends to the flat space
metric sufficiently fast as $r\rightarrow\infty$ (see \cite{Sachs:1962zza} for
further details).
This construction allows one to define asymptotic symmetries and then, quoting
Sachs, to ``separate the kinematics of space-time from the dynamics of the
gravitational field at least at spatial infinity'' \cite{Sachs:1962zza}. It is
worth noting that such a separation can only be done in asymptotically flat
spacetimes, by looking at asymptotic symmetries; otherwise, ``in general
relativity such a separation is usually impossible'' \cite{Sachs:1962zza}.

Asymptotic symmetries are the transformations preserving the asymptotically
flat form of the metric, i.e.  preserving the above properties. These
symmetries form a group: the BMS group. The vector fields generating such
transformations satisfy an equation which is weaker than the Killing equation,
and coincides with the Killing equation as $r\rightarrow\infty$. The BMS group
is defined by the following transformations of the $(u,\theta,\phi)$
coordinates:
\begin{eqnarray}
\lim_{r\rightarrow\infty}\theta'&=&\theta'(\theta,\phi)\,,\nn\\
\lim_{r\rightarrow\infty}\phi'&=&\phi'(\theta,\phi)\,,\nn\\
\lim_{r\rightarrow\infty}u'&=&K(\theta,\phi)[u-\alpha(\theta,\phi)]\,,
\label{BMS0}
\end{eqnarray}
where $\theta'(\theta,\phi)$, $\phi'(\theta,\phi)$ and $K(\theta,\phi)$ are
related by a conformal transformation, i.e.
\begin{equation}
d\theta^{\prime 2}+\sin^2\theta' d\phi^{\prime2}=K^2\left( d\theta^2+
\sin^2\theta d\phi^2\right)\,.
\label{confsphere}
\end{equation}
We follow the convention of Newman and Penrose \cite{Newman:1966ub}, but note
that Ref.~\cite{Goldberg:1966uu} uses a different convention: their $K$ is our
$K^{-1}$. We can consider Eqs.~(\ref{BMS0}) as transformations defined on a
three-dimensional manifold described by $(u,\theta,\phi)$, i.e.  on the future
null hypersurface:
\begin{eqnarray}
\theta'&=&\theta'(\theta,\phi)\,,\nn\\
\phi'&=&\phi'(\theta,\phi)\,,\nn\\
u'&=&K(\theta,\phi)[u-\alpha(\theta,\phi)]\label{BMS}\,.
\end{eqnarray}
The transformations apply to asymptotic quantities defined at future null
infinity, which are obtained by expanding tensors (or Newman-Penrose scalars)
in powers of $1/r$ \cite{newman:891}.  This approach is particularly
meaningful in the study of gravitational-wave emission, since the detector's
location is usually a good approximation to future null infinity.

The BMS group is the semidirect sum of the Lorentz group with the group of
{\it supertranslations}, which are a generalization of translations. The
supertranslations are the transformations (\ref{BMS}) with $\theta'=\theta$,
$\phi'=\phi$ and $K(\theta,\phi)=1$. A useful illustration of their meaning is
given in Ref.~\cite{mccarthy:1837}.  Translations are special cases of
supertranslations, with $\alpha(\theta,\phi)=\epsilon_0+\epsilon_in_i$.  Here
$i=1,2,3$ and $n_i=x_i/r=(\sin\theta\cos\phi,\sin\theta\sin\phi,\cos\theta)$
are director cosines.

Lorentz transformations are transformations of the form (\ref{BMS}) with
$\alpha(\theta,\phi)=0$:
\begin{eqnarray}
\theta'&=&\theta'(\theta,\phi)\,,\nn\\
\phi'&=&\phi'(\theta,\phi)\,,\nn\\
u'&=&K(\theta,\phi)u\,.\label{BMSG}
\end{eqnarray}
Strictly speaking, there is no unique way to identify the Lorentz group as a
subgroup of the BMS group \cite{Geroch1977,Newman:1966ub}. Indeed, if $\{L\}$
are the transformations (\ref{BMSG}) and $T(\alpha)$ is a supertranslation
with fixed $\alpha$, then the transformations
$\{T(\alpha)\,L\,T^{-1}(\alpha)\}$ form another group isomorphic to the
Lorentz group. As far as we know, there is no rigorous proof that our choice
(\ref{BMSG}) of the Lorentz subgroup is more ``natural'' than others, but
there are arguments supporting this choice.  In particular,
Ref.~\cite{Newman:1966ub} suggests that with further, reasonable physical
assumptions, the Lorentz group embedded in the BMS group is restricted to
$\{T(\alpha)\,L\,T^{-1}(\alpha)\}$ with $\alpha=\epsilon_0+\epsilon_in_i$
(ordinary translations); the residual freedom, then, corresponds to an
arbitrariness in the choice of the origin. Therefore it is not restrictive to
assume $\alpha=0$, i.e. to assume that the Lorentz transformations are given
by Eq.~(\ref{BMSG}).
Ordinary rotations are special cases of Lorentz transformations with
$K(\theta,\phi)=1$. As discussed in Appendix \ref{boostapp}, a boost with
(non-relativistic) three-velocity $\vec V=(V_1,V_2,V_3)$ has
$K(\theta,\phi)=1+V_in_i$.

To study Lorentz transformations, and then conformal transformations on the
$2$-sphere, it is useful to introduce the stereographic representation, in
which the coordinates $(\theta\,,\phi)$ on the $2$-sphere are mapped on the
complex plane by introducing a complex variable
\begin{equation}
\zeta\equiv \cot\frac{\theta}{2}\,e^{\ii\phi}\,.
\label{defzeta}
\end{equation}
Lorentz transformations correspond to mappings $\zeta\to \zeta'$ of the form
\begin{equation}
\zeta'=\frac{a\zeta+c}{b\zeta+d}\,,
\label{LorStereo}
\end{equation}
where $a,b,c,d$ are complex numbers and we can assume without loss of
generality that $ad-bc=1$. The interested reader can find a more detailed
discussion in Appendix \ref{stereoapp}.

\section{BMS transformations of gravitational radiation}
\label{BMStrans}

As we discussed above, $l^\mu\equiv g^{\mu\nu}u_{,\nu}$ is a null vector, and
$r$ is an affine parameter of the null geodesic tangent to $l^\mu$. Following
\cite{newman:566,Newman:1966ub,newman:891}, we normalize $r$ by requiring that
$l^\mu=dx^\mu/dr$; furthermore, we take $l^\mu$ to be a vector of the
Newman-Penrose tetrad.  It is important to stress that this is a particular
choice of the tetrad \cite{Lehner:2007ip}: we are fixing the tetrad boost
freedom $l^\mu\rightarrow\lambda l^\mu$, $n^\mu\rightarrow\lambda^{-1} n^\mu$
(and then the normalization of $\Psi_4$, since for a tetrad boost
$\Psi_4\rightarrow\lambda^{-2}\Psi_4$).
Ref.~\cite{newman:891} calls this choice the ``special coordinate system and
associated tetrad'', and Ref.~\cite{Lehner:2007ip} calls it the ``Bondi
tetrad.'' Most of the literature on the asymptotic behavior of Newman-Penrose
quantities and on BMS transformations assumes this tetrad choice. The same
normalization has been used by Teukolsky to study perturbations of Kerr black
holes \cite{Teukolsky:1973ap}. As discussed in
\cite{Lehner:2007ip,Buonanno:2007pf}, recent works dealing with numerical
evolutions of the Einstein equations use a tetrad choice such that
$l_\alpha=u_{,\alpha}/\sqrt{2}$
\cite{Buonanno:2006ui,Brugmann:2008zz,Sperhake:2006cy,Berti:2007fi}.  We do
not follow the latter convention because Eq.~(\ref{psi40s}) below, as derived
in \cite{newman:891}, requires $l_\alpha=u_{,\alpha}$.

\subsection{Weyl scalar and shear}

Gravitational radiation is described by the Weyl scalar $\Psi_4$. The
asymptotic limit of $\Psi_4$ at future null infinity is closely related to the
asymptotic limit of the shear $\sigma$, which is defined in terms of the
Newman-Penrose tetrad as
\begin{equation}
\sigma(u,r,\theta,\phi)=l_{\mu;\nu}m^\mu m^\nu\,.
\end{equation}
Indeed, assuming the special coordinate system and the associated tetrad
defined above, the asymptotic expansion of $\Psi_4$ and $\sigma$ is
\cite{newman:891}
\begin{eqnarray}
\Psi_4(r,u,\theta,\phi))&=&\frac{\Psi_4^0(u,\theta,\phi)}{r}+{\cal O}(r^{-2})\,,
\nonumber\\
\sigma(r,u,\theta,\phi)&=&\frac{\sigma^0(u,\theta,\phi)}{r^2}+{\cal O}(r^{-4})\,,
\end{eqnarray}
where
\begin{equation}
\Psi_4^0=-\frac{\pa^2\bar{\sigma}^0}{\pa u^2}\,,
\label{psi40s}
\end{equation}
and an overbar denotes complex conjugation. To simplify calculations, in the
remainder of this paper we will usually consider SWSHs of spin weight $s=2$
and expand the complex-conjugate Weyl scalar as:
\begin{equation}
  \bar\Psi_4^0=e^{-\ii\omega u}
  \sum_{lm}\,\,_{2}\psi_{lm}\,_2Y_{lm}(\theta,\phi)\,,
  \label{expandcc}
\end{equation}
where we assume a gravitational-wave solution with given frequency $\omega$.
In numerical relativity simulations the Weyl scalar is usually expanded in
SWSHs of spin-weight $s=-2$. The relevant expansion can easily be obtained by
recalling that SWSHs satisfy the property
\begin{equation}
_sY_{lm}(\theta,\phi)=(-1)^{m+s}\,_{-s}\bar Y_{l\,-m}(\theta,\phi)\,,
\label{barY}
\end{equation}
which implies that
\begin{equation}
\Psi^0_4=e^{{\rm i}\omega u}\sum_{lm}\,_{-2}\psi_{lm}\,_{-2}Y_{lm}(\theta,\phi)
\end{equation}
with
\be
\,_{-2}\psi_{lm}\equiv (-1)^m  \,_{2}{\bar \psi}_{l-m}\,.\label{prop0}
\ee

The first derivative of the shear with respect to retarded time,
$\pa\sigma^0/\pa u$, is called the {\em news function}. Its squared modulus
represents the energy density of the gravitational wave: the total energy
emitted by the source (corresponding to its mass loss) is \cite{Sachs:1962wk}
\begin{equation}
E_{\rm tot}=\frac{1}{4\pi}\int\left|\frac{\pa\sigma^0}{\pa u}\right|^2
dud\Omega\,.\label{Etot}
\end{equation}
To understand the relation of $\sigma^0$ with the gravitational-wave
amplitude, let us consider the asymptotic metric in the transverse-traceless
(TT) gauge; following the notation of Teukolsky \cite{Teukolsky:1973ap},
\begin{equation}
\Psi_4=\frac{1}{2}(\ddot h_+-\ii\ddot h_\times)=\frac{\ddot h}{2}\,,
\end{equation}
where we have defined the complex wave amplitude $h$ in terms of the two
polarization modes $h_+$ and $h_\times$ as $h\equiv h_+-\ii h_\times$.
Assuming again a gravitational-wave solution with given frequency,
\begin{equation}
h=\frac{1}{r}\,e^{\ii\omega u}\hat h\,,
\end{equation}
we have
\begin{equation}
\Psi_4=\frac{1}{2}\frac{\pa^2 h}{\pa u^2}\,.\label{ourpsi}
\end{equation}
From Eq.~(\ref{psi40s}) it follows that
\begin{equation}
rh=e^{\ii\omega u}\hat h=-2\bar\sigma^0\,;
\label{sigma0h}
\end{equation}
therefore, $\sigma^0$ corresponds (modulo proportionality constants) to the
wave amplitude $h$.

From Eq.~(\ref{Etot}) we have
\begin{equation}
\frac{d^2E}{dtd\Omega}=\frac{1}{4\pi}
\left|\frac{d^2\sigma_0}{du^2}\right|= \frac{\omega^2r^2}{16\pi}|h|^2
=\frac{\omega^2r^2}{16\pi}(|h_+|^2+|h_\times|^2)\,,
\end{equation}
which is the well-known expression for the energy carried by a gravitational
wave \cite{Misner:1973cw}. Notice that
\begin{equation}
\frac{d^2E}{dtd\Omega}=\frac{r^2|\Psi_4^2|}{4\pi\omega^2}\,,
\end{equation}
which is consistent with Ref.~\cite{Teukolsky:1973ap}.  In
\cite{Buonanno:2006ui,Brugmann:2008zz,Sperhake:2006cy,Berti:2007fi}, where
$l_\alpha=u_{,\alpha}/\sqrt{2}$, the Weyl scalar is $\Psi_4=(\ddot
h_+-\ii\ddot h_\times)$. With their normalization,
$d^2E/dtd\Omega=r^2|\Psi_4^2|/16\pi\omega^2$. The interpretation of the
asymptotic shear as the amplitude of the wave in the TT-gauge is also
discussed in Ref.~\cite{Brown:1996bw}.

Under a BMS transformation the asymptotic shear transforms as
\cite{Sachs:1962wk,Newman:1966ub}
\begin{equation}
\sigma^{0\prime}(u,\theta,\phi)= K^{-1}e^{2\ii\chi}\left(\sigma^0(u,\theta,\phi)
-\frac{1}{2}/\!\!\!\pa^2\alpha\right)\,,\label{BMSshear}
\end{equation}
%
where the operator $/\!\!\!\partial$, acting on a quantity $\eta$ of spin
weight $s$, is defined by
\begin{equation}
/\!\!\!\partial\eta=-(\sin\theta)^s\left(\frac{\partial}{\partial\theta}
+\frac{\ii}{\sin\theta}\frac{\partial}{\partial\phi}\right)
(\sin\theta)^{-s}\eta\,, 
\end{equation} 
and the angle
$\chi$ represents the rotation of the vectors $m^\mu$, $\bar m^\mu$ of
the Newman-Penrose tetrad.  For Lorentz transformations $\alpha=0$, and Eq.
(\ref{BMSshear}) reduces to
\begin{equation}
\sigma^{0\prime}(u,\theta,\phi)=
K^{-1}e^{2\ii\chi(\theta,\phi)}\sigma^0(u,\theta,\phi)\,.
\label{BMSshear1}
\end{equation}
Incidentally, this relation also holds for ordinary translations, since
$/\!\!\!\pa^2\alpha=0$ in this case.  The total radiated energy (\ref{Etot})
transforms as
\begin{equation}
E_{\rm tot}'=\frac{1}{4\pi}\int\left|\frac{\pa\sigma^{0\prime}}{\pa u'}\right|
du'd\Omega'=\frac{1}{4\pi}\int K^{-4}\left|\frac{\pa\sigma^0}{\pa u}\right|
Kdu\,K^2d\Omega
=\frac{1}{4\pi}\int K^{-1}\left|\frac{\pa\sigma^0}{\pa u}\right|
dud\Omega\,,
\end{equation}
with a shift factor $K^{-1}$, as expected. This relation is quite interesting:
in the case of a boost, it describes the beaming of the emitted gravitational
radiation.
%

The transformation (\ref{BMSshear1}) is not yet complete, since
$\sigma^{0\prime}$ is still expressed in the unprimed coordinates
$(u,\theta,\phi)$.  The complete transformation involves the relation between
primed and unprimed coordinates:
\begin{equation}
\sigma^{0\prime}(u',\theta',\phi')=
K^{-1}e^{2\ii\chi(\theta(\theta'),\phi(\phi'))}
\sigma^0(u(u'),\theta(\theta'),\phi(\phi'))\,.
\label{BMSshear2}
\end{equation}
For a gravitational wave with frequency $\omega$ we have
$\sigma^0(u,\theta,\phi)=e^{-\ii\omega u}\hat \sigma^0(\theta,\phi)$, and
since $u'=Ku$,
\begin{equation}
\sigma^{0\prime}(u',\theta',\phi')=
K^{-1}e^{2\ii\chi(\theta(\theta'),\phi(\phi'))}
e^{-\ii\frac{\omega}{K}u'}\hat\sigma^0(\theta(\theta'),\phi(\phi'))\,.
\label{BMSshear3}
\end{equation}
We see that the frequency of the gravitational wave is blueshifted or
redshifted: $\omega\rightarrow\omega'=\omega/K$.

The transformation of the asymptotic Weyl scalar $\bar\Psi_4^0$ (we consider
the complex conjugate $\bar\Psi_4^0$, with spin $2$, instead of $\Psi_4$, with
spin $-2$, to simplify calculations), related to $\sigma^0$ by
Eq.~(\ref{psi40s}), immediately follows:
\begin{equation}
\bar\Psi_4^{0\prime}(u',\theta',\phi')=
K^{-3}e^{2\ii\chi(\theta(\theta'),\phi(\phi'))}
\Psi_4^0(u(u'),\theta(\theta'),\phi(\phi'))\,.
\label{BMSpsi4}
\end{equation}
We now expand the complex-conjugate Weyl scalar in SWSHs of spin weight $s=2$,
as in Eq.~(\ref{expandcc}). Given the transformation of the harmonics
\begin{equation}
_2Y_{l'm'}(\theta'(\theta,\phi),\phi'(\theta,\phi))=
\sum_{lm}\,D_{l'm';lm} \,_2Y_{lm}(\theta,\phi)\,,\label{plainharmstrasf}
\end{equation}
the Lorentz transformed Weyl scalar is
\begin{equation}
\bar\Psi_4^{0\prime}(u',\theta',\phi')=
K^{-3}e^{2\ii\chi(\theta(\theta'),\phi(\phi'))}
e^{-\ii\frac{\omega}{K} u'}
\sum_{l'm'}\,
\,_{2}\psi'_{l'm'}\,_2Y_{l'm'}(\theta',\phi')\,,\\
\label{BMSpsi4transf0}
\end{equation}
with
\begin{equation}
\,_{2}\psi'_{l'm'}=\sum_{lm}\,\,_{2}\psi_{lm}D^{-1}_{lm;l'm'}\,.
\label{BMSpsi4transf}
\end{equation}
In the case of rotations $K=1$ and $D_{lm;l'm'}$ is block diagonal in $l$,
because spherical harmonics with fixed $l$ and $s$ form irreducible
representations of the rotation group. Then the transformation of the Weyl
scalar reduces to
\begin{subequations}
\begin{eqnarray}
\bar\Psi_4^{0\prime}(u',\theta',\phi')&=&
e^{2\ii\chi(\theta(\theta'),\phi(\phi'))}
e^{-\ii\omega u'}
\sum_{lm'}
\,_{2}\psi'_{lm'}\,_2Y_{lm'}(\theta',\phi')\,,\label{rotpsi4transf0}\\
\,_{2}\psi'_{lm'}&=&\sum_{m=-l}^l\, _{2}\psi_{lm}D^{(l)\,-1}_{m;m'}\,,
\label{rotpsi4transf}
\end{eqnarray}
\end{subequations}
where we have defined
\begin{equation}
D_{m m'}^{(l)}\equiv D_{lm;lm'}\,.\label{defDlblock}
\end{equation}
Therefore, the transformation of the multipolar decomposition of the Weyl
scalar under rotations is given by the (inverse) transformation of $s=2$
SWSHs, modulo an overall phase factor.

We now turn to a detailed treatment of the transformation of SWSHs under BMS
transformations. We start with the general case, then specialize to rotations
and non-relativistic boosts. Finally we show that there is a simple relation
between the transformations of the Weyl scalar components in a SWSH expansion,
and the transformations of SWSHs themselves.

\subsection{Transformations of spin-weighted spherical harmonics}

The BMS transformations of SWSHs are discussed in Ref.~\cite{Goldberg:1966uu}.
%
%
%
Given $s$ and $l\ge|s|$, choose arbitrarily an integer $L\ge l$. Then SWSHs
can be written as
\begin{equation}
_sY_{lm}=\sum_{n_1=0}^{L-s}\sum_{n_2=0}^{L+s}\,_sB^{L\,n_1n_2}_{lm}
\,_sZ^L_{n_1n_2}\,,
\label{SWSHexp}
\end{equation}
where $_sB^{L\,n_1n_2}_{lm}$ are constants whose explicit values can be found
in Ref.~\cite{Goldberg:1966uu}, and $_sZ^L_{n_1n_2}$ are properly defined
functions of $\zeta$ and $\bar\zeta$. This equation can be rewritten in a
simpler form:

\begin{equation}
_2Y_{lm}=(-1)^m\sqrt{\frac{2l+1}{4\pi}}
\sqrt{\frac{(l+m)!(l-m)!}{(l+2)!(l-2)!}}
\sum_n(-1)^n
\left(\begin{array}{c}l-2\\n\\\end{array}\right)
\left(\begin{array}{c}l+2\\l-m-n\\\end{array}\right)Z^n_{lm}\,,
\label{sterharmall}
\end{equation}
where $\max(0,-2-m)\leq n\leq l-\max(m,2)$ and we have defined
\begin{equation}
Z^n_{lm}\equiv (\zeta\bar\zeta)^{l-2-n}
\bar\zeta^{2-m}(1+\zeta\bar\zeta)^{-l}\,.\label{Zall}
\end{equation}
This expression is consistent with Eq.~(4.13) of Ref.~\cite{Goldberg:1966uu}
if we correct for an erroneous factor of $(-1)^m$ in their Eq.~(2.12).  For
$l=2$ and $l=3$ the SWSHs reduce to
\begin{subequations}
\begin{eqnarray}
_2Y_{2m}&=&(-1)^m\sqrt{\frac{30}{\pi}}
\frac{1}{\sqrt{(2+m)!(2-m)!}}Z_m\,,\label{sterharml2}\\
_2Y_{3m}&=&(-1)^{m+1}\sqrt{\frac{210}{\pi}}
\left[\frac{1-\delta_{m,3}}{(2-m)!}\sqrt{\frac{(3-m)!}{(3+m)!}}Z^{(+)}_m-
\frac{1-\delta_{m,-3}}{(2+m)!}\sqrt{\frac{(3+m)!}{(3-m)!}}Z^{(-)}_m
\right]\,,\label{sterharml3}
\end{eqnarray}
\label{sterharm}
\end{subequations}
where we have defined
\begin{subequations}
\begin{eqnarray}
Z_m\equiv\,_2Z_{0\,m+2}^2&=&\frac{1}{(1+\zeta\bar\zeta)^2}\bar\zeta^{2-m}=
\left(\sin\frac{\theta}{2}\right)^{2+m}\left(\cos\frac{\theta}{2}\right)^{2-m}
e^{\ii m\phi}e^{-2\ii\phi}\,,\qquad (-2\le m\le2)\,,\\
Z^{(+)}_m\equiv\,_2Z_{1\,m+3}^3&=&\frac{1}{(1+\zeta\bar\zeta)^3}\bar\zeta^{2-m}=
\left(\sin\frac{\theta}{2}\right)^{4+m}\left(\cos\frac{\theta}{2}\right)^{2-m}
e^{\ii m\phi}e^{-2\ii\phi}\,,\qquad (-3\le m\le2)\,,\\
Z^{(-)}_m\equiv\,_2Z_{0\,m+2}^3&=&\frac{\zeta\bar\zeta}{(1+\zeta\bar\zeta)^3}
\bar\zeta^{2-m}=
\left(\sin\frac{\theta}{2}\right)^{2+m}\left(\cos\frac{\theta}{2}\right)^{4-m}
e^{\ii m\phi}e^{-2\ii\phi}\,,\qquad (-2\le m\le3)\,.
\end{eqnarray}
\end{subequations}
The factor $e^{-2\ii\phi}$ is due to different conventions in defining SWSHs,
corresponding to different choices of the Newman-Penrose tetrad (see footnote
6 of \cite{Goldberg:1966uu}).
We have checked that Eqs.~(\ref{sterharmall}) and (\ref{sterharm}) are
consistent with the standard definition of SWSHs in terms of ordinary
spherical harmonics:
\begin{equation}
_2Y_{lm}=\frac{1}{\sqrt{(l-1)l(l+1)(l+2)}}\left(\pa^2_\theta
-\cot\theta\pa_\theta+\frac{m^2}{\sin^2\theta}-\frac{2m}{\sin\theta}
\left(\pa_\theta-\cot\theta\right)\right)Y_{lm}\,,\label{standharm}
\end{equation}
where we have followed the conventions of \cite{Goldberg:1966uu}, adopted also
in recent numerical simulations
\cite{Buonanno:2006ui,Brugmann:2008zz,Sperhake:2006cy,Berti:2007fi}.  The only
difference between Eq.~(\ref{standharm}) and our expressions is a global
factor of $e^{-2\ii\phi}$, which does not depend on $m$ and is therefore
irrelevant. The $s=-2$ harmonics can be obtained by complex conjugation, using
Eq.~(\ref{barY}).

Let us focus for the time being on $l=2$.  In stereographic coordinates (see
Appendix \ref{stereoapp}) a Lorentz transformation maps $\zeta\to\zeta'$ as in
Eq.~(\ref{LorStereo}), and
\begin{equation}
Z'_{m'}=K^2e^{2\ii\chi}\frac{1}{(1+\zeta\bar\zeta)^2}
(\bar a\bar\zeta+\bar c)^{2-m'}(\bar b\bar\zeta+\bar d)^{2+m'}
=K^2e^{2\ii\chi}\sum_{m=-2}^2F_{m'm}Z_m\,,\qquad (-2\le
m\le2)\,,\label{trasfZ}
\end{equation}
where the constants $F_{m'm}$ are given by the expansion coefficients of
polynomials of the form
\begin{equation}
(\bar a\bar\zeta+\bar c)^{2-m'}(\bar b\bar\zeta+\bar d)^{2+m'}
=\sum_{m=-2}^2F_{m'm}\bar\zeta^{2-m}\,,\qquad (-2\le m'\le2)\,.
\label{defF}
\end{equation}
Replacing Eqs.~(\ref{trasfZ}) and (\ref{defF}) in Eq.~(\ref{sterharml2})
we find the transformation of spin-2 SWSHs with $l=2$ and its inverse:
\begin{eqnarray}
_2Y_{2m'}(\theta',\phi')
&=&K^2(\theta,\phi)e^{2\ii\chi(\theta,\phi)}\sum_{m=-2}^2A^{(2)}_{m'm}
\,_2Y_{2m}(\theta,\phi)\,,\label{transfsterl2}\\
_2Y_{2m}(\theta,\phi)&=&K^{-2}e^{-2\ii\chi(\theta',\phi')}
\sum_{m'=-2}^2 A^{(2)\,-1}_{mm'}
\,_2Y_{2m'}(\theta',\phi')\,,\label{transfsterl2inv}
\end{eqnarray}
where
\begin{equation}
A^{(2)}_{m'm}\equiv(-1)^{m+m'}\sqrt{\frac{(2+m)!(2-m)!}{(2+m')!(2-m')!}}F_{m'm}\,.
\label{Ammp}
\end{equation}
The constant coefficients of the inverse matrix $A^{(2)\,-1}_{mm'}$ can be
derived using the inverse Lorentz transformation
\begin{equation}
\zeta=\frac{d\zeta'-c}{-b\zeta'+a}\,,
\label{inverzezeta}
\end{equation}
or the identity
\begin{equation}
\sum_{m=-2}^2A^{(2)}_{m'm}A^{(2)\,-1}_{mm''}=\delta_{m'm''}\,.
\end{equation}
For $l>2$ the transformation of the harmonics is more involved.
Eq.~(\ref{transfsterl2inv}) can easily be generalized only in the case of pure
rotations, due to the decoupling of harmonics with different values of
$l$, but the general Lorentz transformations are more difficult to
work out explicitly.

Below we discuss two special cases in which the formalism simplifies
considerably: pure rotations and non-relativistic boosts with $|V|\ll 1$.

\subsubsection{Transformations under rotations}

Since spin-$s$ SWSHs with a given value of $l$ are a representation of the
rotation group, for pure rotations we get
\begin{equation}
_2Y_{lm'}(\theta',\phi')=\sum_{m=-l}^l D^{(l)}_{m'm}\,_2Y_{lm}(\theta,\phi)\,.
\end{equation}
where $D^{(l)}_{m'm}$ has been defined in Eq.~(\ref{defDlblock}).
Since $K=1$ for pure rotations, Eq.~(\ref{transfsterl2}) gives
\begin{equation}
D^{(2)}_{m'm}=e^{2\ii\chi}A^{(2)}_{m'm}\,,
\label{D2A2}
\end{equation}
with $A^{(2)}_{m'm}$ a constant matrix. As we discuss below, this property
holds for all $l$'s: $D^{(l)}_{m'm}=e^{2\ii\chi}A^{(l)}_{m'm}$, and therefore
\begin{equation}
_2Y_{lm'}(\theta',\phi')=e^{2\ii\chi}\sum_mA^{(l)}_{m'm}
\,_2Y_{lm}(\theta,\phi)\,,
\label{rotl}
\end{equation}
with $A^{(l)}_{m'm}$ a constant matrix.

By appropriately choosing the primed frame, a general rotation can be reduced
to a rotation mapping any vector (such as the total angular momentum) to the
$z'$-axis:
\begin{equation}
(J_x,J_y,J_z)=(J\sin\theta_0\cos\phi_0,J\sin\theta_0\sin\phi_0,
J\cos\theta_0)~~\rightarrow~~(0,0,J)\,.
\end{equation}
As discussed in Appendix \ref{stereoapp}, the stereographic representation of
such a rotation has the form
\begin{equation}
a=c_0 e^{-\ii\frac{\phi_0}{2}}\,,\quad
b=-s_0 e^{-\ii\frac{\phi_0}{2}}\,,\quad
c=s_0 e^{\ii\frac{\phi_0}{2}}\,,\quad
d=c_0 e^{\ii\frac{\phi_0}{2}}\,,
\label{rotlaw}
\end{equation}
%
where we have used the short-hand notation
\begin{equation}
c_0\equiv\cos(\theta_0/2)\,,\qquad
s_0\equiv\sin(\theta_0/2)\,.
\label{c0s0}
\end{equation}
The corresponding transformation matrix for $l=2$ SWSHs with $s=2$,
$A_{m'm}^{(2)}$, is
\begin{eqnarray}
&&e^{\ii m\phi_0}\,A_{m'm}^{(2)}= \left( \begin {array}{ccccc} c_0^4&-2c_0^3s_0
&\sqrt{6}c_0^2s_0^2&-2c_0s_0^3&s_0^4\\
2{c_0}^{3}s_0&c_0^2(1-4s_0^2)&\sqrt {6}s_0 c_0(1-2c_0^2)&4s_0^2(c_0^2-1/4)&-2c_0
s_0^3\\
\sqrt {6}c_0^2 s_0^2&-\sqrt{6}s_0c_0(1-2c_0^2)&1-6c_0^2s_0^2&
\sqrt{6}s_0c_0(1-2c_0^2)&\sqrt{6}c_0^2s_0^2\\
2c_0s_0^3&4s_0^2(c_0^2-1/4)&-\sqrt{6}s_0c_0(1-2c_0^2)&c_0^2
(1-4s_0^2)&-2{s_0}{c_0}^{3}\\
s_0^4&2c_0s_0^3&\sqrt{6}c_0^2s_0^2&2{c_0}^{3}s_0&{c_0}^{4}\\
\end {array} \right)\,.
\label{Ammprot}
\end{eqnarray}
Here and below, whenever we write the transformation in matrix form, different
rows correspond to increasing values of $m'(=-l,\dots,l)$, and different
columns correspond to increasing values of $m(=-l,\dots,l)$.

As a useful special case, consider the transformation of the $l=2$ harmonics
under a counterclockwise rotation by $\pi/2$ around the $x_2$-axis. In
stereographic coordinates this transformation reduces to
\begin{equation}
\zeta'=\frac{\frac{\zeta}{\sqrt{2}}+\frac{1}{\sqrt{2}}}{-\frac{\zeta}{\sqrt{2}}
+\frac{1}{\sqrt{2}}}\,,
\label{pi2}
\end{equation}
so Eq.~(\ref{defF}) becomes
\begin{equation}
\frac{1}{4}(-1)^{m'}(\bar\zeta-1)^{2+m'}(\bar\zeta+1)^{2-m'}
=F_{m'm}\bar\zeta^{2-m}\,.
\end{equation}
The corresponding coefficients $A^{(2)}_{m'm}$ are
\begin{equation}
\label{transforml2}
A^{(2)}_{m'm}=\frac{1}{4}\left(\begin{array}{ccccc}
1&-2&\sqrt{6}&-2&1\\2&-2&0&2&-2\\\sqrt{6}&0&-2&0&\sqrt{6}\\
2&2&0&-2&-2\\1&2&\sqrt{6}&2&1\\\end{array}\right)\,.
\end{equation}
%

The transformation under rotations of spin-2 SWSHs with $l>2$ is
sligthly more involved. Under a rotation, the functions $Z^n_{lm}$
defined in Eq.~(\ref{Zall}) transform to
\begin{equation}
Z^{\prime\,n'}_{lm'}=\frac{e^{2\ii\chi}}{(1+\zeta\bar\zeta)^l}
(\bar a\bar\zeta+\bar c)^{l-m'-n'}(\bar b\bar\zeta+\bar d)^{2+m'+n'}
(a\zeta+c)^{l-2-n'}(b\zeta+d)^{n'}
=e^{2\ii\chi}\sum_{mn}F_{m'm}^{n'n}Z^n_{lm}\,,
\label{expZpgen}
\end{equation}
which defines (implicitly) the constant coefficients $F_{m'm}^{n'n}$. In
Eq.~(\ref{expZpgen}) the indices $n,n'$ are defined in the ranges
\begin{equation}
\max(0,-2-m)\leq n\leq l-\max(m,2)\,,~~~~~~~
\max(0,-2-m')\leq n'\leq l-\max(m',2)\,,
\end{equation}
respectively.

Correspondingly, the spherical harmonics transform to
\begin{eqnarray}
_2Y'_{lm'}&=&(-1)^{m'}\sqrt{\frac{2l+1}{4\pi}}
\sqrt{\frac{(l+m')!(l-m')!}{(l+2)!(l-2)!}}
\sum_{n'}(-1)^{n'}
\left(\begin{array}{c}l-2\\n'\\\end{array}\right)
\left(\begin{array}{c}l+2\\l-m'-n'\\\end{array}\right)
\left(e^{2\ii\chi}\sum_{mn}F_{m'm}^{n'n}Z^n_{lm}\right)\nonumber\\
&=&e^{2\ii\chi}(-1)^m\sqrt{\frac{2l+1}{4\pi}}
\sqrt{\frac{(l+m)!(l-m)!}{(l+2)!(l-2)!}}
\sum_n(-1)^n
\left(\begin{array}{c}l-2\\n\\\end{array}\right)
\left(\begin{array}{c}l+2\\l-m-n\\\end{array}\right)
A^{(l)}_{m'm}
Z^n_{lm}\,.
\end{eqnarray}
where we defined
\begin{equation}
A^{(l)}_{m'm}
\equiv
\sum_{n'}(-1)^{m+m'+n+n'}
\sqrt{\frac{(l+m')!(l-m')!}{(l+m)!(l-m)!}}
\frac{\left(\begin{array}{c}l-2\\n'\\\end{array}\right)
\left(\begin{array}{c}l+2\\l-m'-n'\\\end{array}\right)}{
\left(\begin{array}{c}l-2\\n\\\end{array}\right)
\left(\begin{array}{c}l+2\\l-m-n\\\end{array}\right)}
F_{m'm}^{n'n}\,.\label{defAlmpm}
\end{equation}
%
Remarkably, the right-hand side of Eq.~(\ref{defAlmpm}) does not seem
to depend on the index $n$. This property implies Eq.~(\ref{rotl}),
which shows explicitly that SWSHs with fixed $l$ and $s$ form an
irreducible representation of the rotation group, as expected from
group-theoretical considerations.  We do not have an analytic proof
that $A^{(l)}_{m'm}$ is independent of $n$. However we checked this
property for $l\leq 10$ and all values of $m,m'$ (which include all
cases of interest for gravitational wave detection), and we expect it
to hold in general.

The coefficients $A^{(l)}_{m'm}$ can explicitly be computed case-by-case
from Eq.~(\ref{defAlmpm}). Note that the phase factor $e^{2\ii\chi}$ is
independent of $l$. As we will show, this phase factor cancels out in
the transformation of the Weyl scalars.

For reference, let us evaluate explicitly the components of the rotation
matrix $A^{(l)}_{m'm}$ for $l=3$.  For a general rotation by angles
$(\theta_0\,,\phi_0)$ the transformation matrix $A_{m'm}^{(3)}$ is
\begin{eqnarray}
&&e^{im\phi_0} \,A_{m'm}^{(3)}= \left( \begin {array}{ccccccc}
\\c_0^6&-\sqrt{6}c_0^5s_0&\sqrt{15}c_0^4s_0^2&
-2\sqrt{5}s_0^3c_0^3&\dots&\dots&\dots\\
\sqrt{6}c_0^5s_0&c_0^4(1-6s_0^2)&-\sqrt{5/2}c_0^3s_0(2-6s_0^2)&
\sqrt{30}s_0^2c_0^2(1-2s_0^2)&
\dots&\dots&\dots\\
\sqrt{15}c_0^4s_0^2&\sqrt{5/2}c_0^3s_0(2-6s_0^2)&c_0^2(1-10s_0^2+15s_0^4)
&-\left[2c_0s_0+5\sin(3\theta_0)\right]/16&\dots&\dots&\dots\\
2\sqrt{5}c_0^3s_0^3&\sqrt{30}\,c_0^2s_0^2(1-2s_0^2)&2\sqrt{3}c_0s_0
(1-5s_0^2+5s_0^4)&\left[3\cos\theta_0+5\cos(3\theta_0)\right]/8&\dots&\dots&\dots\\
\sqrt{15}c_0^2s_0^4&\sqrt{5/2}c_0s_0^3(4-6s_0^2)&s_0^2(6-20s_0^2+15s_0^4)&
\sqrt{3}\left[\sin\theta_0+5\sin(3\theta_0)\right]/16&\dots&\dots&\dots\\
\sqrt{6}c_0s_0^5&s_0^4(5-6s_0^2)&\sqrt{5/2}c_0s_0^3(4-6s_0^2)&
\sqrt{30}s_0^2c_0^2(1-2s_0^2)&\dots&\dots&\dots\\
s_0^6&\sqrt{6}c_0s_0^5&\sqrt{15}c_0^2s_0^4&2\sqrt{5}s_0^3c_0^3&\dots&\dots&\dots\\
\end {array} \right)\,,\nn\\
&&\label{l3general}
\end{eqnarray}
where again we have used the short-hand notation defined in Eq.~(\ref{c0s0}).
The missing components can be obtained with the help of the general relation
\be
A_{m'm}^{(l)}=(-1)^{m+m'}
\bar{A}_{-m'-m}^{(l)}\,.\label{genrel1}\\
\ee
%
In the special case of a $\pi/2$ counterclockwise rotation around the
$x_2$-axis, from these relations and Eq.~(\ref{pi2}) we get
\begin{eqnarray}
\label{l3pi2}
&&A_{m'm}^{(3)}= \frac{1}{8}
\left( \begin {array}{ccccccc} \\1&-\sqrt{6}&\sqrt{15}&
-2\sqrt{5}&\sqrt{15}&-\sqrt{6}&1\\
\sqrt{6}&-4&\sqrt{10}&0&-\sqrt{10}&4&-\sqrt{6}\\
\sqrt{15}&-\sqrt{10}&-1&2\sqrt{3}&-1&-\sqrt{10}&\sqrt{15}\\
2\sqrt{5}&0&-2\sqrt{3}&0&2\sqrt{3}&0&-2\sqrt{5}\\
\sqrt{15}&\sqrt{10}&-1&-2\sqrt{3}&-1&\sqrt{10}&\sqrt{15}\\
\sqrt{6}&4&\sqrt{10}&0&\sqrt{10}&-4&-\sqrt{6}\\
1&\sqrt{6}&\sqrt{15}&
2\sqrt{5}&\sqrt{15}&\sqrt{6}&1\\
\end {array} \right)\,.
\end{eqnarray}

\subsubsection{Transformations under non-relativistic boosts}

In the case of boosts, Appendix \ref{stereoapp} shows that the conformal
factor is $K=1+V_in_i$ to linear order in the boost velocity. At the same
order, the transformation in the stereographic representation is described by
the parameters
%
\begin{eqnarray}
a=1-\frac{V_3}{2}\,,\quad
b=-\frac{V_-}{2}\,,\quad
c=-\frac{V_+}{2}\,,\quad
d=1+\frac{V_3}{2}\,,
\label{boosts}
\end{eqnarray}
%
where we have defined $V_{\pm}\equiv V_1\pm iV_2$, and $V_i\ll1$ is the boost
velocity along the $x_i$-axis (see Appendix \ref{stereoapp}). The components
$A^{(2)}_{m'm}$ can now be determined from Eqs.~(\ref{defF}) and (\ref{Ammp}).
To first order in $V_i$ they read
\begin{eqnarray}
\label{boost}
&&A^{(2)}_{m'm}= \left( \begin {array}{ccccc}
1-2V_3&V_+&0&0&0\\
V_-&1-V_3&\sqrt{3/2}\,V_+&0&0\\
0&\sqrt{3/2}\,V_-&1&\sqrt{3/2}\,V_+&0\\
0&0&\sqrt{3/2}\,V_-&1+V_3&V_+\\
0&0&0&V_-&1+2V_3\\\end {array} \right)\,.
\end{eqnarray}
We stress that, although only $l=2$ harmonics formally appear in the expansion
(\ref{transfsterl2}), for a boost there is no true decoupling in $l$, since
the conformal factor $K$ has a non-trivial dependence on $\theta$ and $\phi$.
Harmonics with $l>2$ would appear if we expanded in constant coefficients, as
in Eq.~(\ref{plainharmstrasf}).

\subsection{Transformations of the Weyl scalar}

In order to find the Lorentz transformation of the Weyl scalar $\Psi_4$, as
given in Eqs.~(\ref{BMSpsi4transf0}) and (\ref{BMSpsi4transf}),
%
%
we need to determine the inverse transformation of spin-2 SWSHs, i.e. the
coefficients $D_{lm;l'm'}^{-1}$. We will do so in the following.

\subsubsection{Transformations under rotations}

Under rotations, using Eqs.~(\ref{rotpsi4transf0}) and (\ref{rotl}),
Eq.~(\ref{BMSpsi4transf0}) reduces to
\begin{subequations}
\begin{eqnarray}
\bar\Psi_4^{0\prime}(u',\theta',\phi')&=&
e^{-\ii\omega u}
\sum_{lm'}
\,_{2}\psi'_{lm'}\,_2Y_{lm'}(\theta',\phi')\,, \label{equatepsi4}\\
\,_{2}\psi'_{lm'}&=&\sum_{m} \,_{2}\psi_{lm}A^{(l)\,-1}_{mm'}\,.
\label{Cprime}
\end{eqnarray}
\end{subequations}
It is worth noting that the phase factors $e^{\pm 2\ii\chi}$ cancel, and the
transformation under rotations of the coefficients of the multipolar expansion
of $\bar\Psi_4$ coincides with the inverse transformation of spin-2 SWSHs.

The transformation laws for the standard decomposition of the Weyl
scalar $\Psi_4$ in $s=-2$ harmonics can be obtained with the help of
relations (\ref{barY}), (\ref{prop0}).  However, a simpler relation
emerges if one uses Eq.~(\ref{genrel1}).  Using the complex conjugate
of Eqs.~(\ref{Cprime}) and (\ref{genrel1}) one can show that
\begin{subequations}
\begin{eqnarray}
\Psi_4^{0\prime}(u',\theta',\phi')&=&
e^{\ii\omega u}
\sum_{lm'}\,_{-2}\psi'_{lm'} \,_{-2}Y_{lm'}(\theta',\phi')\,,\\
\,_{-2}\psi'_{lm'}&=&\sum_{m} \,_{-2}\psi_{lm} A^{(l)\,-1}_{mm'}\,.
\label{Cprimeminus2}
\end{eqnarray}
\end{subequations}
This means that the coefficients of $s=-2$ and $s=+2$ transform with the same
representation. Some applications of Eq.~(\ref{Cprimeminus2}) to numerical
relativity simulations of black hole mergers are discussed in Section
\ref{numerical}.

\subsubsection{Transformations under boosts}\label{boostmixing}

The case of pure boosts is more complex than a pure rotation. The conformal
factor $K$ is now a function of $\theta$ and $\phi$, so it is not trivial to
extract the transformation law of the coefficients $_2\psi_{lm}$ from
Eq.~(\ref{BMSpsi4transf0}).

Quite interestingly, the factor $K(\theta,\phi)$ couples harmonics with
different values of $l$. To see how this coupling comes about and how boosts
affect the mode decomposition, let us consider a gravitational wave which in
the original frame consists of an $(l=2,m=0)$ mode only. If we move to a frame
which is boosted by $V_3$ along the $x_3$-axis and we neglect the phase factor
$e^{-\ii\frac{\omega}{K}u'}$ (e.g. by considering wavefronts for which
$u'=0$), from Eq.~(\ref{BMSpsi4transf0}) we get
\begin{equation}
\bar\Psi_4^{0\prime}(u',\theta',\phi')\approx (1-5V_3\cos\theta')\,_{2}\psi_{20}\sum_{m'}
A^{(2)\,-1}_{0m'}\,_2Y_{2m'}(\theta',\phi')\,,
\label{Psi4allboost}
\end{equation}
where we expanded $K^{-5}=(1+V_3\cos\theta)^{-5}\simeq 1-5V_3\cos\theta$. The
inverse transformation matrix, ignoring second-order terms, is
\begin{eqnarray}
&&A^{(2)\,-1}_{mm'}= \left( \begin {array}{ccccc}
1-2V_3&0&0&0&0\\
0&1+V_3&0&0&0\\
0&0&1&0&0\\
0&0&0&1-V_3&0\\
0&0&0&0&1-2V_3\\\end {array} \right)\,.
\end{eqnarray}
If we equate Eq. (\ref{Psi4allboost}) to Eq.~(\ref{equatepsi4})
%
%
we get
\begin{equation}
\sum_{lm}\,_{2}\psi'_{lm} \,{_2}Y_{lm}(\theta',\phi')=
\,_{2}\psi_{20}(1-5V_3\cos\theta')\,_2Y_{20}(\theta',\phi')
=\,_{2}\psi_{20}
\left[_2Y_{20}(\theta',\phi')
-\frac{5}{\sqrt{7}}V_3\,_2Y_{30}(\theta',\phi')\right]\,,
\end{equation}
where we used the property
\begin{equation}
\cos\theta'\,_2Y_{20}(\theta',\phi')=
\frac{1}{\sqrt{7}}\,_2Y_{30}(\theta',\phi')\,.
\end{equation}
Therefore, the boost produces an $l=3$ component: the nonvanishing components
of the multipolar expansion are
\begin{equation}
\,_{2}\psi'_{20}=\,_{2}\psi_{20}\,,\qquad 
\,_{2}\psi'_{30}=-\frac{5}{\sqrt{7}}V_3\,_{2}\psi_{20}\,.
\label{boostmix}
\end{equation}
An $l=2$ waveform is seen as distorted in the boosted frame, with
contributions also from $l=3$. If we took into account higher powers of $V$
we would find that higher multipoles would also be excited.

Some considerations on the physical relevance of this coupling are in order.
For unequal-mass spinning black hole binaries, the final black hole can be
kicked with velocities as large as $4000~{\rm km}/{\rm s}$
\cite{Campanelli:2007ew,Campanelli:2007cga,Gonzalez:2007hi,Dain:2008ck}.
In the center-of-mass frame, numerical simulations (and possibly also
post-Newtonian expansions) will see an $l=3$ component due entirely to the
fact that the black hole is boosted. The $l=3$ component can be as large as
$V_3\sim 10^{-2}$ times the $l=2$ component.
Boost-induced multipole coupling will certainly be relevant in small
impact-parameter collisions of ultrarelativistic black holes: in this case the
velocity is a sizeable fraction of the speed of light, and the energy transfer
between modes can have important consequences. A detailed study of this
problem will be reported elsewhere.

\section{Application to numerical simulations and black hole perturbation theory}
\label{numerical}

In this section we apply the tools developed in the rest of the paper to
numerical simulations of binary black-hole mergers and to a classical problem
in black-hole perturbation theory. We examine in detail two scenarios. 

In the first scenario, we consider two possible physical situations: the
head-on collision of equal-mass, non-spinning black holes in numerical
relativity, and the infall of a point particle into a Schwarzschild black hole
in perturbation theory. We study the collision in two different frames: (1) a
frame where the collision (or infall) occurs along the $z$-axis, making
maximal use of the symmetries of the system, and (2) a frame where the
collision occurs in the equatorial plane (along the $x$-axis, for the
equal-mass case). The two frames are obviously related by a $\pi/2$ rotation:
$\theta\to \theta'=\theta-\pi/2$. Our main purpose is to show in a relatively
simple but non-trivial setting that our formalism is consistent with the
predictions of numerical simulations and of black-hole perturbation theory.

In the second scenario we consider a more general case: the merger of
unequal-mass, spinning black holes in a so-called ``superkick'' configuration
\cite{Campanelli:2007ew,Campanelli:2007cga,Gonzalez:2007hi,Dain:2008ck}. This
configuration is general enough to provide a good case study for generic,
unequal-mass mergers of spinning black holes. 
Our main goal in this exercise is to stress the importance, in the context of
generic mergers, of relating two reference frames: the ``simulation frame''
(where the binary initially orbits on the $x-y$ plane, so that the orbital
angular momentum initially points in the $z$-direction) and the ``remnant
frame'' (defined as the frame where the final black hole's spin is aligned
with the $z$-axis). To a good level of approximation, if we ignore the
(relatively small) boost, the two frames are related by a rotation. We suggest
a simple method to estimate the angles $(\theta_0\,,\phi_0)$ describing this
rotation.  
Then we argue that the analysis of merger waveforms is more ``natural'' in the
remnant frame.  More specifically, we show that (i) the merger radiation, as
seen from the $z$-axis, is circularly polarized in the remnant frame, but not
in the simulation frame, and (ii) ringdown frequencies provide reasonable
estimates of the final black-hole spin in the remnant frame, but not in the
simulation frame.

\subsection{Head-on collisions of non-spinning black holes}
\label{headon}

Let us start by discussing how rotations affect the description of two
``simple,'' highly symmetric processes leading to the generation of
gravitational waves: the head-on collisions of equal-mass, non-spinning black
holes in numerical relativity, and the infall of a point particle into a
Schwarzschild black hole in perturbation theory.

\subsubsection{Equal mass head-on collisions in numerical relativity}

\begin{figure}[hbt]
\centering
\includegraphics[width=6.8cm,angle=-90]{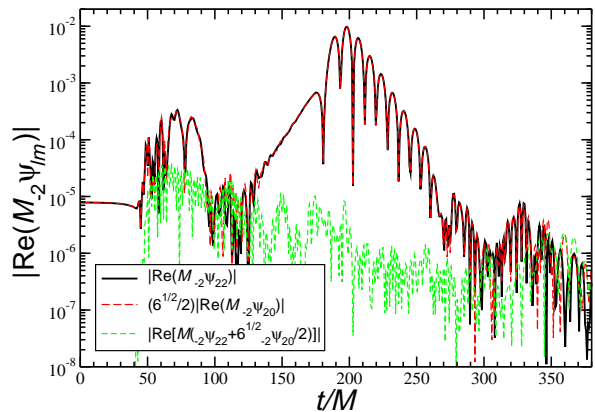}
\caption{Check of the transformation law for the $l=2$ modes of the head-on
  collisions studied in \cite{Sperhake:2007gu}. To better quantify the error,
  the short-dashed (green) line shows $|{\rm
    Re}[M(_{-2}\psi_{22}+\sqrt{6}_{-2}\psi_{20}/2)]|$. This quantity should be
  zero, and indeed it is typically much smaller than the corresponding
  amplitudes, with the exception of the early inspiral and late ringdown
  phases.
%
%
}
\label{fig:headon}
\end{figure}

For an equal mass head-on collision of non-spinning holes along the $z$-axis,
by symmetry, only modes with $m=0$ are excited (see \cite{Sperhake:2006cy} and
references therein). In Ref.~\cite{Sperhake:2007gu} we studied the transition
from circular inspirals to head-on collisions by considering a sequence of
eccentric inspirals. Initially the black holes were placed on the $x$-axis.
Along the sequence, we fixed the system's binding energy. The (equal and
opposite) linear momenta on the two holes were directed along the $y$-axis,
and gradually reduced from the value $P=P_{\rm circ}$ required to produce a
quasi-circular inspiral down to $P=0$. In the limit $P=0$ the black holes
collide head-on along the $x$-axis. By analyzing the resulting radiation we
observed that it contains components with $(l=2,m=\pm 2)$ and $(l=2,m=0)$, and
that these components have roughly comparable amplitude (see top left panel in
Fig.~2 of \cite{Sperhake:2007gu}). Now we are in a position to predict
quantitatively the relative amplitude of these components. In the frame where
the holes collide along the $z$-axis, we should only have a component with
$(l=2,m=0)$. By colliding the holes along the $x$-axis, we are effectively
performing a $\pi/2$ rotation in the $\theta$-direction.  But then from
Eq.~(\ref{transforml2}) we immediately see that the $(l=2,m=2)$ and
$(l=2,m=0)$ wave amplitudes should be proportional to each other, with a
proportionality factor of $-\sqrt{6}/2$. 
In Fig.~\ref{fig:headon} we show that this prediction is in excellent
agreement with the numerical simulations. The error (short-dashed, green
curve) is orders of magnitude below the wave signal except for those times
when the GWs are so weak that $\Psi_4$ is dominated by numerical noise; that
is, during the early inspiral, once the junk radiation has left the system,
and in the late ringdown stages.
Ref.~\cite{Palenzuela:2006wp} independently derived a similar result in the
context of head-on collisions of boson stars.


\subsubsection{Infall of particles in black-hole perturbation theory}\label{infall}

The previous discussion referred to equal-mass black hole collisions. The
limit where one black hole is much smaller than the other ($M_1\ll M_2$) is
also of great interest, since it can be described within black hole
perturbation theory. In this framework the small black hole is considered as a
perturbation in the geometry of the large black hole, and one can expand
Einstein's equations to first order in the parameter $M_1/M_2$. The resulting
system of equations can be further reduced to a single ordinary differential
equation with a source term ${\cal L}$. The functional form of the source term
depends of course on the frame chosen to do the calculation. In particularly
symmetric situations (e.g. for the radial infall of a small body into a more
massive black hole) the source term has a simple analytical form. For this
reason, the radial infall of a point particle into a black hole is a good
testing ground for the formalism we developed in this paper. We consider the
source term in two frames: one where the infall occurs along the symmetry axis
(so that only modes with $m=0$ are non-vanishing) and a frame rotated by
$\pi/2$, so that the collision occurs in the equatorial plane.

Details of the formalism are discussed in Appendix \ref{particleapp}. Here we
only report the main result: for particles falling along the axis and in the
equatorial plane the source terms are given by
Eqs.~(\ref{sourceaxisl2})-(\ref{explicitS1a0}) and
(\ref{sourceeq20})-(\ref{sourceeq31}), respectively. Since we work in linear
perturbation theory the waveforms are proportional to $\cal L$, and therefore
we infer that
\begin{equation}
\frac{_{-2}\psi_{22}^{\rm equator}}{_{-2}\psi_{20}^{\rm equator}}
=-\frac{\sqrt{6}}{2}\,,\qquad 
\frac{_{-2}\psi_{20}^{\rm axis}}{_{-2} \psi_{20}^{\rm equator}}
=2\,,\qquad
\frac{_{-2}\psi_{3\pm3}^{\rm equator}}{_{-2}\psi_{3\pm1}^{\rm equator}}
=-\sqrt{\frac{5}{3}}\,,\qquad 
\frac{_{-2}\psi_{30}^{\rm axis}}{_{-2} \psi_{33}^{\rm equator}}
=\frac{4}{\sqrt{5}}\,.
\end{equation}
These proportionality relations are perfectly consistent with the
transformation matrices (\ref{transforml2}) and (\ref{l3pi2}).

\subsection{Generic rotation angles: unequal-mass ``superkick'' runs}
\label{genericrot}

As a non-trivial practical application of the transformation laws
(\ref{Ammprot}) and (\ref{Cprimeminus2}), in this section we study the
inspiral and merger of unequal-mass, initially spinning black holes.  Our
study is a rather striking demonstration of the importance of choosing the
``right'' reference frame to analyze the radiation produced by a merger event.

Our case-study is the ``sk4'' run described in Ref.~\cite{Berti:2007nw}. Our
black-hole binary has mass ratio $q\equiv M_1/M_2=4$. The Kerr parameters of
both holes have the same magnitude $|j_1|=|j_2|=0.727$, but opposite
directions, and lie on the orbital plane at the beginning of the simulation:
this is sometimes referred to as a ``superkick'' configuration.

Consider the dimensionless spin vector of the remnant black hole ${\cal
  J}=({\cal J}_x,{\cal J}_y,{\cal J}_z)$.  Define ${\cal J}_{xy}=({\cal
  J}_x^2+{\cal J}_y^2)^{1/2}$, and let ${\cal J}=({\cal J}_x^2+{\cal
  J}_y^2+{\cal J}_z^2)^{1/2}$ be the modulus of the final spin. Simple
flat-space geometry suggests that a rough but reasonable estimate of the
rotation angles relating the simulation frame to the remnant frame should be
\begin{equation}
\theta_0=\arcsin \left({\cal J}_{xy}/{\cal J}\right)\,,\qquad
\phi_0=\arccos \left({\cal J}_x/{\cal J}_{xy}\right)\,.
\end{equation}
For our sk4 run, the final spin can be obtained from angular momentum balance
(see Appendix \ref{radiatedj} for details). The final spin magnitude estimated
in this way is ${\cal J}=0.642$, and the individual spin components are ${\cal
  J}_x=0.424$, ${\cal J}_y=0.036$, ${\cal J}_z=0.480$. The system experiences
a significant realignment: the estimated rotation angles are
$\theta_0=41.5^{\circ}$ and $\phi_0=4.8^{\circ}$.  Now we can use this
estimate of the rotation angles, together with relations (\ref{Ammprot}) and
(\ref{Cprimeminus2}), to transform the ``simulation frame'' waveforms to the
``remnant frame'' waveforms.

\begin{figure}[hbt]
\centering
\includegraphics[width=6.8cm,angle=-90]{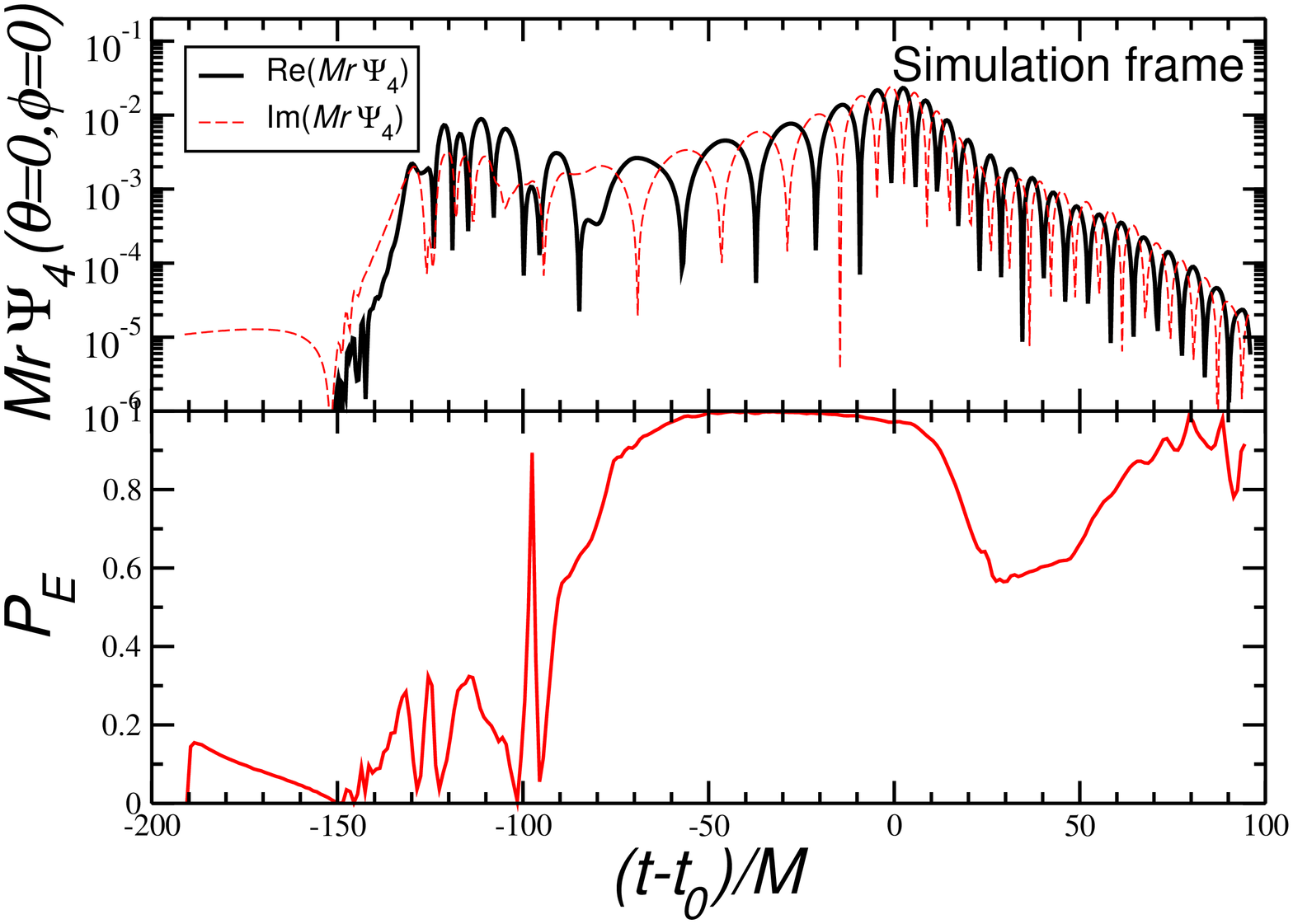}
\includegraphics[width=6.8cm,angle=-90]{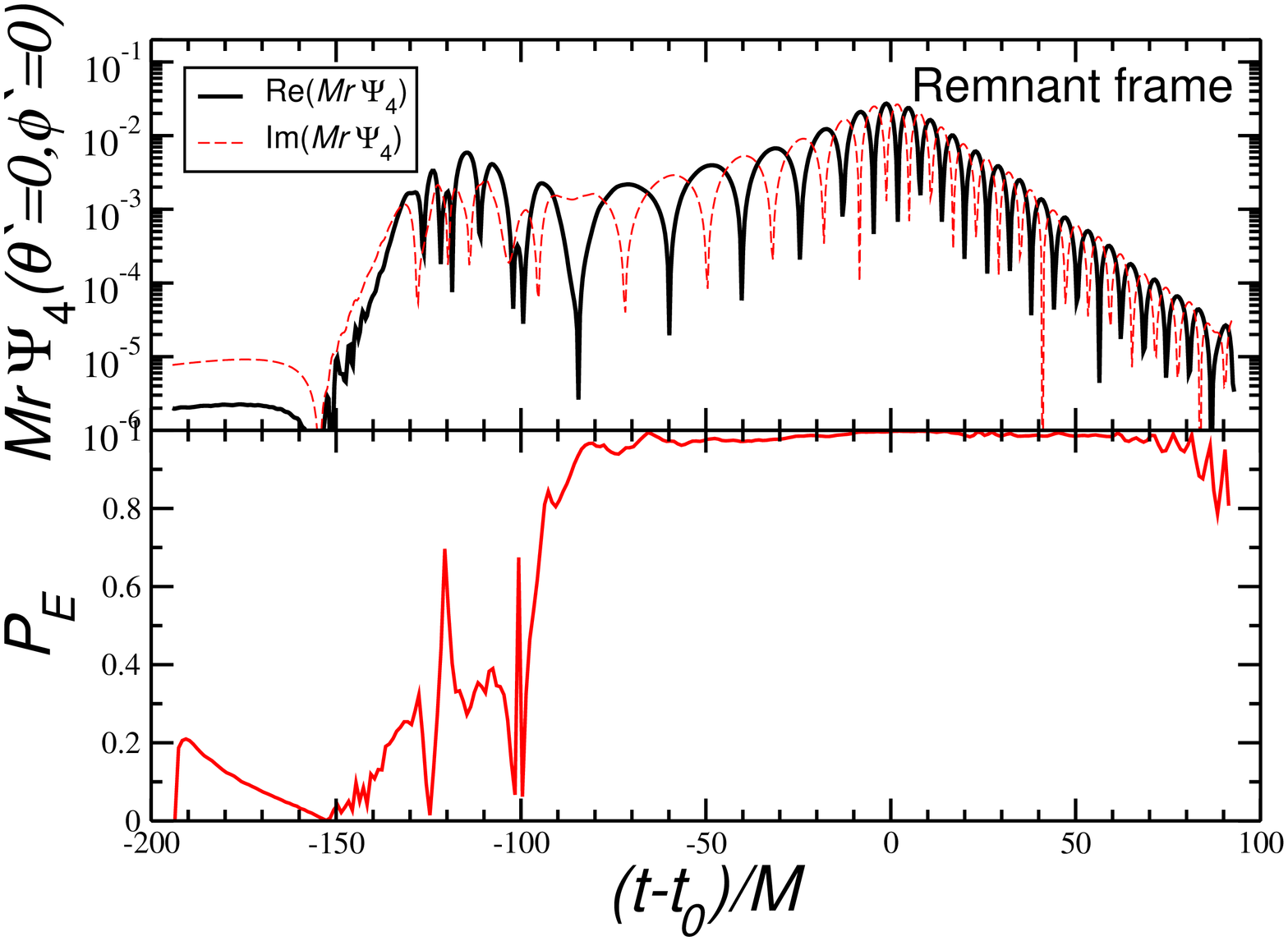}
\caption{Top: Quadrupolar ($l=2$) component of $|{\rm Re}(Mr\,\Psi_{4})|$ and
  $|{\rm Im}(Mr\,\Psi_{4})|$ as seen by an observer located on the $z$-axis,
  for the sk4 run of Ref.~\cite{Berti:2007nw}. Bottom: degree of elliptic
  polarization of the waveforms (compare Fig.~28 in
  Ref.~\cite{Berti:2007fi}). The left panel refers to the simulation frame,
  the right panel to the remnant frame.}
\label{fig:pol}
\end{figure}

Our results are summarized in Fig.~\ref{fig:pol}. The upper left panel of
Fig.~\ref{fig:pol} shows the dominant, quadrupolar ($l=2$) component of
$\Psi_4$, as seen by an observer located on the $z$-axis in the simulation
frame. To obtain this quantity we take the sum
\be
\sum_{l=2} \sum_{m=-2,0,2} \,_{-2}\psi_{lm}(t,r)\,_{-2}Y_{lm}(\theta,\phi)\,,
\ee
of all $l=2$ components with different values of $m$ (i.e. $m=2$, $m=0$ and
$m=-2$), and we evaluate it for $\theta=\phi=0$. The upper right panel plots
the quadrupolar ($l=2$) component of $\Psi_4$, as seen by an observer on the
$z$-axis in the remnant frame. Again, this quantity is obtained by taking the
sum of all $l=2$ components in the new frame, as prescribed by the
transformation laws for $l=2$. In both plots the time axis is shifted by
$t_0$, the time corresponding to the peak in the waveform's amplitude, which
is also a rough measure for the formation of a common apparent horizon and the
beginning of ringdown (see e.g.~\cite{Berti:2007fi}).

Some simplification in the waveform is clearly seen in the remnant frame,
especially in the late (ringdown) phase. In the remnant frame, the ringdown
part of the radiation is well described by a single, "clean" damped
exponential, and the envelope of the maxima is well fitted by a straight
line. In the simulation frame the envelope of the maxima is an oscillating
function.  This happens because the ringdown radiation is effectively a sum of
components with the same $l(=2)$ but different values of $m$, and this
produces beatings between different quasinormal modes.

An alternative measure of the improvement produced by using the remnant frame
is given by the degree of elliptic polarization $P_E$, shown in the bottom
panels of Fig.~\ref{fig:pol}. This quantity was defined and discussed in
Appendix D of Ref.~\cite{Berti:2007fi}. It has the property that $P_E=1$ for
circularly polarized waves, and $P_E=0$ for linear polarization.
Fig.~\ref{fig:pol} shows that the ringdown signal is {\em not} circularly
polarized as seen from the $z$-axis in the simulation frame, but it is indeed
circularly polarized (to a very good approximation) in the remnant frame.
This is a rather impressive demonstration that the merger-ringdown waveforms
look simpler in the remnant frame. On physical grounds this makes perfect
sense, since the remnant frame (unlike the simulation frame) is well adapted
to the final hole. For generic precessing binaries, it is still possible that
switching to the remnant frame does not simplify the early inspiral
waveform. A detailed analysis of longer, generic inspiral simulations will be
a topic for future research.

It was observed a long time ago that the polarization state of the waves is a
useful discriminant between alternative theories of gravity
\cite{Eardley:1973br,Eardley:1974nw}.  More recently, it has been suggested
that LISA could place interesting bounds on the amplitude birefringence (i.e.
the polarization dependent amplification, or attenuation, of the wave
amplitude with distance) predicted by Chern-Simons gravity
\cite{Alexander:2007kv}. The fact that the merger-ringdown waveform is
circularly polarized in the remnant frame may have useful applications to test
general relativity using strong-field observations of merging binaries.

\begin{figure}[hbt]
\centering
\includegraphics[width=6.8cm,angle=-90]{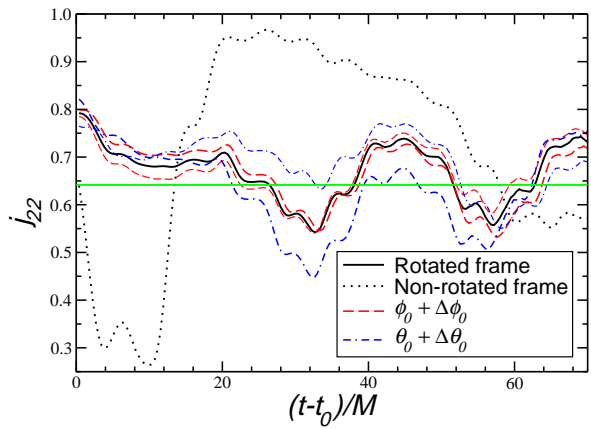}
\caption{Final angular momentum from the $(l=2\,,m=2)$ mode in the remnant
  frame (solid line) and in the simulation frame (dotted line). Dashed thick
  (thin) red lines: we arbitrarily increase (decrease) $\phi_0$ by $\Delta
  \phi_0=4^{\circ}$.  Dash-dash-dotted thick (thin) blue lines: we arbitrarily
  increase (decrease) $\theta_0$ by $\Delta \theta_0=8^{\circ}$. The
  horizontal line represents the estimate of the final spin (${\cal J}=0.642$)
  obtained from balance arguments.}
\label{fig:spin}
\end{figure}

As an additional demonstration of the usefulness of the remnant frame it is
useful to look at the ringdown radiation. We use a Prony analysis of the
$l=m=2$ ringdown mode to extract the quality factor of the quasinormal ringing
\cite{Berti:2005ys}. From the quality factor we compute the Kerr parameter of
the final hole (see \cite{Berti:2007fi} and \cite{Berti:2007dg} for further
details), comparing this estimate with the prediction from angular momentum
balance arguments.  Our results are shown in Fig.~\ref{fig:spin}, where we
show the spin estimated by considering different portions of the ringdown
signal, starting at time $t>t_0$. In the remnant frame, the final spin
estimated from the ringdown signal is much closer to the value estimated from
balance arguments.  The waveform in the simulation frame contains a mixture of
various modes of comparable amplitude and different damping times, and this
produces a large error in the estimation of the final spin. 

As an exercise, in Fig.~\ref{fig:spin} we also show the final spin estimated
in different frames which are ``reasonably close'' to the remnant frame,
considering frames that deviate from our ``best guess'' for the simulation
frame by small angles. For illustration we plot spin estimates obtained
assuming $\theta_0=41.5^{\circ}\pm \Delta \theta_0$ or $\phi_0=4.8^{\circ}\pm
\Delta \phi_0$, where $\Delta \phi_0=4^{\circ}$ and $\Delta
\theta_0=8^{\circ}$. This shows that the good results we obtained in the
remnant frame are not accidental: for example, adding (or subtracting) $\Delta
\theta_0=8^{\circ}$ to our ``best estimate'' for the rotation angle $\theta_0$
seems to produce a systematic underestimate (or overestimate) of the final
spin, as obtained from angular momentum balance arguments. Systematic errors
in our numerical simulations are admittedly quite large, but the significant
bias observed for ``incorrect'' values of $\Delta \theta_0$ is an indication
that frames deviating from the remnant frame by small amounts of rotation can
yield poor agreement with the final spin estimated from balance arguments.

\section{Conclusions and outlook}
\label{conclusions}

We have provided explicit, ready-to-use formulae to transform multipolar
components of gravitational radiation under rotations and non-relativistic
boosts. For generic spinning binaries our results indicate that gravitational
waveforms are simplest in the ``remnant frame'' (see Figs.~\ref{fig:pol} and
\ref{fig:spin}). We provide a practical recipe to identify this frame, and we
suggest that all results from multipolar analyses of merger simulations should
be expressed in the remnant frame to simplify comparisons. Our work will be
useful to standardize file formats \cite{Brown:2007jx} and for the injection
of numerical waveforms in the data analysis of gravitational-wave detectors
\cite{ninja}. Other interesting applications will be (1) an analysis of the
multipolar coupling induced by large boosts in the collision of
ultrarelativistic black holes, and (2) a study of the possibility to determine
the spin orientation of a binary merger remnant using gravitational-wave
observations on Earth and in space. Both topics are currently under
investigation.

\section*{Acknowledgments}
We are grateful to Curt Cutler, Luis Lehner and Nico Yunes for useful
comments on an earlier draft of the manuscript. We also thank the
anonymous referee for his constructive criticism. This work was
supported in part by DFG grant SFB/Transregio~7 ``Gravitational Wave
Astronomy''. V.C.'s work was partially funded by Funda\c c\~ao para a
Ci\^encia e Tecnologia (FCT) - Portugal through projects
PTDC/FIS/64175/2006 and POCI/FP/81915/2007. We thank the DEISA
Consortium (co-funded by the EU, FP6 project 508830), for support
within the DEISA Extreme Computing Initiative (\url{www.deisa.org}).
Computations were performed at LRZ Munich, the Doppler and Kepler
clusters at the Institute of Theoretical Physics of the University of
Jena and the Milipeia cluster at the Center for Computational Physics
(CFC) in Coimbra.  E.B.'s research was supported by an appointment to
the NASA Postdoctoral Program at the Jet Propulsion Laboratory,
California Institute of Technology, administered by Oak Ridge
Associated Universities through a contract with NASA. Government
sponsorship acknowledged.

\appendix

\section{Boosts as BMS transformations}
\label{boostapp}

Transformations of the form (\ref{confsphere}) and (\ref{BMSG})
are conformal transformations of the $(\theta,\phi)$-sphere into itself, with
conformal factor $K$. They correspond to Lorentz transformations of spacetime;
indeed, the conformal group $SL(2,C)$ and the Lorentz group $SO(3,1)$ are
isomorphic, and any Lorentz transformation on spacetime has, on the sphere,
the effect of a conformal transformation (see
\cite{Newman:1966ub,Sachs:1962zza,Goldberg:1966uu,Carmeli:2000af}).

In this Appendix we show this explicitly for the case of a non-relativistic
boost with three-velocity $V_i$ such that $|V|\ll 1$. We also assume that,
since we are interested in asymptotic transformations, $|u|/r\ll1$. We
introduce polar coordinates such that
$x_i=rn_i=r(\sin\theta\cos\phi,\sin\theta\sin\phi,\cos\theta)$, and neglect
terms of order ${\cal O}(V^2)$ and of order ${\cal O}(|u|/r)$. Then under a
boost
\begin{eqnarray}
x'_i&=&x_i-V_it\,,\nn\\
t'&=&t-V_ix_i\,.
\end{eqnarray}
In polar coordinates we have
\begin{eqnarray}
r'&=&\sqrt{x'_ix'_i}
=r-tV_in_i\,,\nn\\
t'&=&t-rV_in_i\,.
\end{eqnarray}
It follows that, to linear order in $V$,
\begin{equation}
u'=t'-r'=u(1+V_in_i)=Ku\,,\label{boostu}
\end{equation}
with $K(\theta,\phi)=1+V_in_i$.
This is our main result. Neglecting terms of order ${\cal O}(|u|/r)$,
\begin{equation}
r'=r\left(1-\frac{u+r}{r}V_in_i\right)=K^{-1}r\,.
\label{boostr}
\end{equation}
The transformation of the director cosines, i.e. of $(\theta,\phi)$, can be
found as follows:
\begin{equation}
x'_i=r'n'_i=(r-tV_jn_j)n'_i=x_i-V_it=rn_i-V_it\,,
\end{equation}
therefore
\begin{equation}
n'_i=\left(1+\frac{t}{r}V_jn_j\right)n_i-\frac{t}{r}V_i=
\left(1+\frac{u+r}{r}V_jn_j\right)n_i-\frac{u+r}{r}V_i
=(1+V_jn_j)n_i-V_i=Kn_i-V_i\,.\label{boostn}
\end{equation}
These are well-defined director cosines: indeed, to linear order in $V$,
\begin{equation}
n'_in'_i=(Kn_i-V_i)(Kn_i-V_i)=K^2-2n_iV_i
=1\,.
\end{equation}
Using the equality $2Kn_iV_i=2(1+V_in_i)n_iV_i=2n_iV_i+{\cal O}(V^2)$ and
consistently neglecting terms of ${\cal O}(V^2)$, it can be shown that the
metric on the two-sphere changes as follows:
\begin{equation}
d\theta^{\prime2}+\sin^2\theta' d\phi^{\prime2}=dn'_idn'_i=
K^2dn_idn_i=K^2\left(d\theta^2+\sin^2\theta d\phi^2\right)\,.
\end{equation}
The boosts, then, act on $(u,\theta,\phi)$ as in Eqs.~(\ref{confsphere}) and
(\ref{BMSG}).  Notice that the flat spacetime metric does not change under a
boost, i.e.
\begin{equation}
ds^2=-du^2-dudr+r^2dn_idn_i=-du^{\prime\,2}-2du'dr'+r^{\prime\,2} dn_i'dn_i'\,,
\end{equation}
but this is not obvious by replacing (\ref{boostu}), (\ref{boostr}) and
(\ref{boostn}) in the above relation. The reason is that we have neglected
terms with $u/r\ll 1$. As a qualitative check of the expression $K=1+V_in_i$,
let us consider the case in which the observers are boosted towards the
source. Then $V_in_i<0$, and $K=1+V_in_i<1$. Therefore, the observed
gravitational-wave frequency changes as $\omega'=\omega/K>\omega$, i.e. we see
a blueshifted wave, as expected.

We have shown that a boost with velocity $\vec V$ such that $|V|\ll 1$ has
conformal factor $K=1+V_in_i$. It affects the transformation of the asymptotic
Weyl scalar $\Psi_4$, Eq.~(\ref{BMSpsi4}), by terms linear in $V$.  We also
have to consider the effect of the change in the coordinates, which is of
order $V$ as can be seen from (\ref{boostu}), (\ref{boostr}) and
(\ref{boostn}), and is harder to derive explicitly. It is derived in a
particular case, in Section \ref{boostmixing}.  Notice that, if the black hole
kick determines both a boost and a translation, the translation only appears
in the change of coordinates, and not in the transformation (\ref{BMSpsi4}).
A translation is expected to give a contribution of order $|\delta\vec x|/r$,
which is extremely small far away from the source: for the purpose of
detection, the boost is the most relevant transformation.

\section{The stereographic representation}
\label{stereoapp}

\begin{figure}[htb]
\includegraphics[width=8cm]{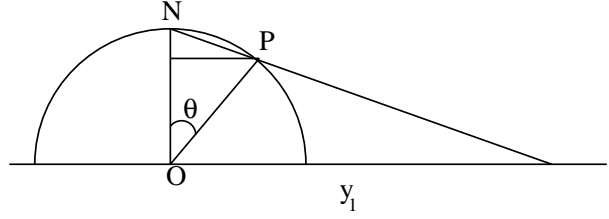}
\caption{Stereographic projection of a sphere onto the plane.
\label{fig:stereo}}
\end{figure}

Conformal transformations of the sphere (which correspond to asymptotic
Lorentz transformations of spacetime) can be expressed as conformal
transformations on the complex plane obtained by projecting the sphere from
the north pole to the equatorial plane $(y_1,y_2)$. From the geometry in
Fig.~\ref{fig:stereo}, defining a complex variable $\zeta=y_1+\ii y_2$ we have

%
%
%
%
\begin{equation}
\zeta=\cot\frac{\theta}{2}e^{\ii\phi}\,.\label{defzeta2}
\end{equation}
The 2-sphere metric can be written as
\begin{equation}
d\theta^2+\sin^2\theta d\phi^2=4(1+\zeta\bar\zeta)^{-2}d\zeta d\bar\zeta\,.
\label{stereog}
\end{equation}
The director cosines $n_i=(\sin\theta\cos\phi,\sin\theta\sin\phi,
\cos\theta)$ are related to $\zeta$ by
%
\begin{equation}
n_1+\ii n_2=
\frac{2\zeta}{\zeta\bar\zeta+1}\,,\qquad
n_3=
\frac{\zeta\bar\zeta-1}{\zeta\bar\zeta+1}\,,
\end{equation}
%
and by the inverse relation, $\zeta
=(n_1+\ii n_2)/(1-n_3)$.
An $SL(2,C)$ transformation
\begin{equation}
A=\left(\begin{array}{cc}a&b\\c&d
\end{array}\right)\,,
\end{equation}
maps $\zeta$ to $\zeta'$ as in Eq.~(\ref{LorStereo}),
%
%
with $ad-bc=1$. Notice that if $ad-bc\neq0$ we can rescale the four elements
without changing the transformation, in order to have $ad-bc=1$. This can be
shown to be a conformal transformation, i.e. the 2-sphere metric changes as in
Eq.~(\ref{confsphere}), which in stereographic coordinates means
\begin{equation}
4(1+\zeta'{\bar\zeta}')^{-2}d\zeta'd{\bar\zeta}'=K^2\,
4(1+\zeta\bar\zeta)^{-2}d\zeta d\bar\zeta\,.\label{confster}
\end{equation}
%
A straightforward calculation shows that
\begin{equation}
K^{-1}=\frac{(a\zeta+c)(\bar a\bar\zeta+\bar c)
+(b\zeta+d)(\bar b\bar\zeta+\bar d)}{\zeta\bar\zeta+1}\,.
\label{defKm1}
\end{equation}
The phase $\chi$ appearing in Eqs.~(\ref{BMSshear}) and (\ref{BMSshear1})
is given by (see Eq.~(3.12) in \cite{Newman:1966ub})
\begin{equation}
e^{-2\ii\chi}=\frac{\pa\zeta'/\pa\zeta}{\pa{\bar\zeta'}/{\pa\bar\zeta}}
=\frac{(\bar b\bar\zeta+\bar d)^2}{(b\zeta+d)^2}\,.
\end{equation}

\subsection{Rotations in stereographic coordinates}

The main purpose of this Section is to derive the specialization of
(\ref{LorStereo}) to a rotation by $(\theta_0,\phi_0)$, i.e.
Eq.~(\ref{rotlaw}), and its special case Eq.~(\ref{pi2}) for rotations by an
angle $\theta_0=\pi/2$. Rotations are $SL(2,C)$ transformations for which
$K=1$. Using Eq.~(\ref{defKm1}) this implies
\begin{equation}
(a\zeta+c)(\bar a\bar\zeta+\bar c) +(b\zeta+d)(\bar b
\bar\zeta+\bar d)=\zeta\bar\zeta+1\,,
\end{equation}
and therefore
\begin{eqnarray}
|a|^2+|b|^2&=&1\,,\nn\\
|c|^2+|d|^2&=&1\,,\nn\\
a\bar c+b\bar d&=&0\,.\label{relrot}
\end{eqnarray}
But $ad-bc=1$, so $A\in SU(2)$ and
\begin{equation}
a=\bar d\,,\qquad b=-\bar c\,.
\end{equation}
Let us first consider, for simplicity, a rotation of $\pi/2$ around the
$x_2$-axis; it maps $x_2\rightarrow x_2$, $x_1\rightarrow x_3$ and
$x_3\rightarrow-x_1$, i.e.
\begin{eqnarray}
\sin\theta'\cos\phi'&=&-\cos\theta\,,\nn\\
\sin\theta'\sin\phi'&=&\sin\theta\sin\phi\,,\nn\\
\cos\theta'&=&\sin\theta\cos\phi\,.\label{rotx2}
\end{eqnarray}
In particular, the plane $\phi=\phi'=0$ is left unchanged, and
$\theta'=\theta-\frac{\pi}{2}$.
The corresponding function $\zeta'(\zeta)$ maps (in the plane $\phi=0$) $-1$
in $0$, $0$ in $1$ and $1$ in $\infty$, therefore it must have the form
\begin{equation}
\zeta'=\frac{a\zeta+c}{b\zeta+d}=\frac{\alpha(\zeta+1)}{\alpha(-\zeta+1)}\,,
\end{equation}
and imposing $ad-bc=1$ this gives Eq.~(\ref{pi2}).
%
%
As expected, this transformation satisfies the relations (\ref{relrot}), thus
$K=1$.

By appropriately choosing the primed frame, a general rotation maps a vector
with generic orientation to the $z'$-axis:
\begin{equation}
(J_x,J_y,J_z)=(J\sin\theta_0\cos\phi_0,J\sin\theta_0\sin\phi_0,
J\cos\theta_0)~~\rightarrow~~(0,0,J)\,.
\end{equation}
Therefore, it maps $\zeta_0=\cot(\theta_0/2)e^{\ii\phi_0/2}$ to
$\zeta'=\infty$. Imposing (\ref{relrot}) (i.e. $K=1$) yields
\begin{equation}
\zeta'=\frac{\cos\frac{\theta_0}{2}e^{-\ii\frac{\phi_0}{2}}\zeta
+\sin\frac{\theta_0}{2}e^{\ii\frac{\phi_0}{2}}}{-
\sin\frac{\theta_0}{2}e^{-\ii\frac{\phi_0}{2}}+
\cos\frac{\theta_0}{2}e^{\ii\frac{\phi_0}{2}}}\,,\label{stergenrot}
\end{equation}
i.e. Eq.~(\ref{rotlaw}).

\subsection{Boosts in stereographic coordinates}

Let us now express boosts in stereographic coordinates. As shown in Appendix
\ref{boostapp}, the conformal factor of a boost with velocity $V_i$ such that
$|V|\ll 1$ is $K=1+V_in_i$. We define
\begin{equation}
V_\pm=V_1\pm\ii V_2\,,
\end{equation}
thus
\begin{equation}
V_in_i=V_+\frac{n_1-\ii n_2}{2}+V_-\frac{n_1+\ii n_2}{2}+V_3n_3
=\frac{1}{\zeta\bar\zeta+1}\left[V_+\bar\zeta+V_-\zeta+V_3(\zeta\bar\zeta-1)
\right]\,.
\end{equation}
Therefore, to linear order in $V$,
\begin{eqnarray}
K&=&1+\frac{1}{\zeta\bar\zeta+1}
\left[V_+\bar\zeta+V_-\zeta+V_3(\zeta\bar\zeta-1)\right]\,,\\
K^{-1}&=&1-\frac{1}{\zeta\bar\zeta+1}
\left[V_+\bar\zeta+V_-\zeta+V_3(\zeta\bar\zeta-1)\right]\,.\label{Km1}
\end{eqnarray}
Now we will work out the expression of a boost as an $SL(2,C)$ transformation.
A four-vector $x^\mu$ in spacetime can be written as an $SL(2,C)$ matrix $X$:
\begin{equation}
X=\sigma_\mu x_\mu=
\left(\begin{array}{cc}x_0+x_3&x_1-\ii x_2\\x_1+\ii x_2&x_0-x_3\\
\end{array}\right)
\end{equation}
where $\sigma_0$ is the identity matrix and $\sigma_i$ ($i=1,2,3$) are
the ordinary Pauli matrices. 
A boost in the $x_1$ direction is represented by the matrix
\begin{equation}
A=e^{-\frac{\sigma_1\omega}{2}}=\cosh\frac{\omega}{2}-\sigma_1
\sinh\frac{\omega}{2}=\left(\begin{array}{cc}\cosh\frac{\omega}{2}
&-\sinh\frac{\omega}{2}\\-\sinh\frac{\omega}{2}&\cosh\frac{\omega}{2}
\end{array}\right)\,,
\end{equation}
where the velocity of the boost is $V=\tanh\omega$. Indeed, the transformed
vector is
\begin{equation}
AXA^\dagger=\left(\begin{array}{cc}
x_0\cosh\omega-x_1\sinh\omega+x_3&
-x_0\sinh\omega+x_1\cosh\omega-\ii x_2\\
-x_0\sinh\omega+x_1\cosh\omega+\ii x_2&
x_0\cosh\omega-x_1\sinh\omega-x_3\\
\end{array}\right)\,.
\end{equation}
where a dagger denotes Hermitian conjugation. For a boost in a generic
direction, defining three parameters $\omega_i$ ($i=1,2,3$) we have
\begin{equation}
A=e^{-\frac{\sigma_i\omega_i}{2}}=\cosh\frac{\omega_i}{2}
-\sigma_i\sinh\frac{\omega_i}{2}\,.
\end{equation}
To linear order $\omega=V+{\cal O}(V^2)$, and
\begin{equation}
A\simeq 1-\sigma_i\frac{V_i}{2}=
\left(\begin{array}{cc}1-\frac{V_3}{2}&-\frac{V_-}{2}\\
-\frac{V_+}{2}&1+\frac{V_3}{2}
\end{array}\right)=\left(\begin{array}{cc}a&b\\c&d
\end{array}\right)\,,
\end{equation}
which is the result quoted in Eq.~(\ref{boosts}).  Such a boost maps
$\zeta=\cot\frac{\theta}{2}e^{\ii\phi}$ into
\begin{equation}
\zeta'=\frac{a\zeta+c}{b\zeta+d}=\frac{\left(1-\frac{V_3}{2}\right)\zeta
-\frac{V_+}{2}}{1+\frac{V_3}{2}-\frac{V_-}{2}\zeta}
\simeq\zeta\left(1-\frac{V_3}{2}+\frac{V_-}{2}\right)
-\frac{V_3}{2} -\frac{V_-}{2}\,,
\end{equation}
and
\begin{equation}
e^{\ii\chi}=\frac{b\zeta+d}{\bar b\bar\zeta+\bar d}=
\frac{1+\frac{V_3}{2}-\frac{V_-\zeta}{2}}{1+\frac{V_3}{2}-
\frac{V_+\bar\zeta}{2}}\simeq1+\frac{V_+\bar\zeta+V_-\zeta}{2}
=1+\ii\cot\frac{\theta}{2}[V_2\cos\phi-V_1\sin\phi]\,.
\end{equation}
As a check of Eq.~(\ref{boosts}) it is easy to verify that the inverse conformal
factor (\ref{defKm1}), when expanded to linear order in $V$, agrees with
Eq.~(\ref{Km1}).

%

\section{Calculation of the radiated angular momentum in numerical evolutions
  of black-hole binaries}
\label{radiatedj}

All numerical black-hole binary evolutions in this paper have been performed
using the {\sc Lean} code \cite{Sperhake:2006cy}. We compute the energy,
linear and angular momentum radiated during merger using the Newman-Penrose
scalar $\Psi_4$. The calculation of $\Psi_4$, as implemented in the {\sc Lean}
code, is described in detail in Appendix~C of Ref.~\cite{Sperhake:2006cy}.
Here we assume $\Psi_4$ as given on an extraction sphere of constant
coordinate radius $r_{\rm ex}$.

The radiated energy, the $(x\,,y\,,z)$ components of the radiated linear
momentum and the $z$ component of the radiated angular momentum are then
given, for example, by Eqs.~(22)-(24) of Ref.~\cite{Campanelli:1998jv}.  For
the simplest binary black hole configurations these quantities encapsulate the
entire information about the radiated momenta, because the $x$ and $y$
components of $J_{\rm rad}$ are known to vanish from symmetry arguments.
However, for more general classes of spinning binaries (such as those
considered in this paper) we can no longer neglect these components.  The
calculation of $J_{x, \rm rad}$ and $J_{y, \rm rad}$ from $\Psi_4$ has
recently been discussed in \cite{Lousto:2007mh,Ruiz:2007yx}. Here we follow
Eqs.~(2.24,~2.25) of \cite{Ruiz:2007yx} and calculate the radiated angular
momentum from
\begin{equation}
  \frac{dJ_{i, \rm rad}}{dt} = -\lim_{r\rightarrow \infty}
      \frac{r^2}{16\pi} \mathrm{Re}
      \oint
        \hat{J}_i \left[
        \int_{-\infty}^t
          \left(
          \int_{-\infty}^{\tilde{t}}
            \Psi_4
          d\bar{t} \right)
        d\tilde{t} \right]
        \left(
        \int_{-\infty}^{t}
          \Psi_4
        d\tilde{t} \right)
      d\Omega,
\end{equation}
where
%
\begin{eqnarray}
  \sin \theta \hat{J}_x &=& -\sin\theta \sin\phi \partial_{\theta}
       - \cos\theta \cos \phi \partial_{\phi} + is\cos\phi, \nn\\
  \sin \theta \hat{J}_y &=& \sin\theta \cos \phi \partial_{\theta}
       - \cos \theta \sin \phi \partial_{\phi} + is\cos \phi, \nn\\
  \hat{J}_z &=& \partial_{\phi},
\end{eqnarray}
%
and the spin weight $s=-2$.

\begin{figure}[hbt]
\centering \includegraphics[width=6.8cm,angle=-90]{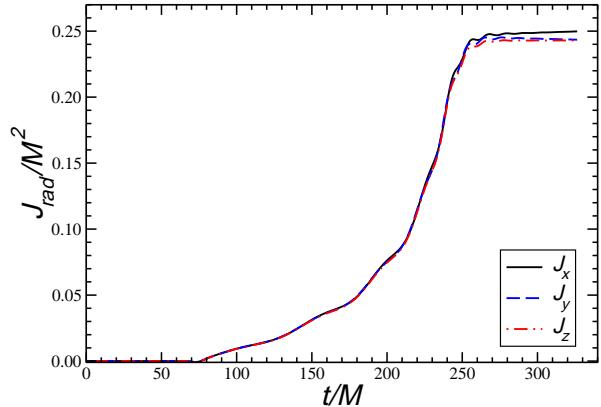}
\caption{Radiated angular momentum as a function of time from the R1
  simulation with the orbital angular momentum aligned with the $x$, $y$ and
  $z$ axis (see text).}
\label{fig:testJxyz}
\end{figure}

As a test of the numerical implementation of these expressions we followed the
suggestion of Sec.~III of Ref.~\cite{Lousto:2007mh}.  We evolved the binary
configuration labelled ``R1'' in table I of Ref.~\cite{Baker:2006yw} three
times, the orbital angular momentum $L$ being aligned with the $z$, $y$ and
$x$ axis, but always using the $z$ axis as reference for the wave extraction.
We use the grid setup of Table I in \cite{Sperhake:2006cy} adapted to the
orientation of $L$, using a comparatively low resolution $h=1/32$ for these
test simulations.  The resulting radiated angular momentum is shown in
Fig.~\ref{fig:testJxyz}.  The radiated angular momenta in the three
simulations agree within a few percent, which for these low resolution runs is
less than the uncertainties resulting from the discretization.


\section{Radial infall of particles into non-rotating black holes}
\label{particleapp}
Small perturbations of the Kerr geometry are usually described using the
Teukolsky formalism \cite{Teukolsky:1972le,Teukolsky:1973ap}, but some
practical applications require a modification of the equations. The
Sasaki-Nakamura formalism
\cite{Sasaki:1981sx,Nakamura:1981kk,Sasaki:1981kj,1984PThPh..71...79K,1983PhLA...96..335K,1983PhLA...99...37K,1981PhLA...87...85S,Hughes:1999bq,Kennefick:1998ab}
is one such modification that is very useful (among other applications) to
deal with the infall of point particles into Kerr black holes. For simplicity,
here we consider a highly relativistic point particle falling into a
non-rotating black hole, since in this case the equations simplify
considerably \cite{Cardoso:2002ay,Cardoso:2002jr}.  Our purpose is to check
the formalism developed in the text by considering the same problem in two
different reference frames. In the first case, we consider the radial infall
of a particle along the $z$-axis, making full use of the symmetries of the
problem. In the second case, we deliberately let the particle fall in along a
radial trajectory on the equatorial plane, destroying the symmetry of the
problem. The point of this Appendix is to derive the source terms of the
Sasaki-Nakamura equation in these two configurations; using such expression,
in Section \ref{infall} we show that the transformation properties derived in
this paper are perfectly compatible with the transformation laws of the source
term of the Sasaki-Nakamura equation.

Details about the Teukolsky formalism may be found in the original literature
\cite{Teukolsky:1972le,Teukolsky:1973ap}, and also in
\cite{1975LNP....44.....B}.
%
%
%
Working in the Kinnersley tetrad, the equations for the Newman-Penrose
quantities decouple and separate, giving rise to the Teukolsky equation for a
radial function $R$ \cite{Teukolsky:1972le,Teukolsky:1973ap}:
\begin{eqnarray}
\Delta^{-s}\frac{d}{dr}\left(\Delta^{s+1}\frac{d}{dr} {R}\right)
-\, _sV \, R =-\, T \,, \label{teukolskyequation}
\label{teukolskyequationexplanation}
\end{eqnarray}
where we defined $\Delta \equiv r^2-2Mr \,$, and $M$ is the mass of the black
hole.  The quantity $T$ is a source term due to the stress-energy of the
particle, and the potential
\begin{eqnarray}
 _sV=\frac{-K^2}{\Delta}+ \frac{\ii sK \Delta'}{\Delta}
-\frac{2\ii sK'}{\Delta}-2\ii sK'+\, _s\lambda\,.
\end{eqnarray}
We work in the frequency domain, assuming a time dependence of the form
$e^{-\ii \omega t}$. The function $K\equiv r^2\omega$, and primes denote
derivatives with respect to $r$. The quantity $s$ is the spin weight (or
helicity) of the field under consideration (see for example
\cite{newman:566}), and $m$ is an azimuthal quantum number. We are interested
in gravitational perturbations, which have spin-weight $s=-2$. For
non-rotating black holes the separation constant ${_s}\lambda=l(l+1)-s(s+1)$.
The source term $T$ appearing in Eq.~(\ref{teukolskyequation}) depends on
the stress-energy tensor of the perturbation under consideration (see e.g.
\cite{1975LNP....44.....B}).  For a test particle falling in from infinity
with zero velocity $T \sim r^{7/2}$, and the potential ${_s}V$ is always
long-ranged. This implies that when we try to solve the Teukolsky equation
(\ref{teukolskyequation}) numerically we will run into problems. Usually the
numerical solution is accomplished by integrating the source term multiplied
by certain homogeneous solutions. At infinity the integral will not be
well-defined: the source term blows up there, and because the potential is
long range, the homogeneous solution will also blow up at infinity.  We can
remedy this long-range nature of the potential, and at the same time
regularize the source term, by using the Sasaki-Nakamura formalism. After some
manipulation, the Teukolsky equation (\ref{teukolskyequation}) may be cast in
the Sasaki-Nakamura form
\cite{Sasaki:1981sx,Nakamura:1981kk,Sasaki:1981kj,1984PThPh..71...79K,1983PhLA...96..335K,1983PhLA...99...37K,1981PhLA...87...85S}:
\begin{equation}
\frac{d^2}{dr_*^2} {X(\omega,r)}- {\cal F}\frac{d}{dr_*}{X(\omega,r)} -
{\cal U} X(\omega,r) = {\cal L} \,.
\label{sn}
\end{equation}
The functions ${\cal F}$ and ${\cal U}$ can be found in the original
literature
\cite{Sasaki:1981sx,Nakamura:1981kk,Sasaki:1981kj,1984PThPh..71...79K,1983PhLA...96..335K,1983PhLA...99...37K,1981PhLA...87...85S}.
The source term ${\cal L}$ is given by
\begin{equation}
{\cal L}=\frac{\gamma_0\Delta}{r^5}W e^{-\ii \int K/\Delta dr_*}\,,
\label{sourceterm}
\end{equation}
and $W$ satisfies
\begin{equation}
W''=-\frac{r^2}{\Delta}T \exp\left[\ii \int \frac{K}{\Delta} dr_*\right]\,.
\label{W}
\end{equation}
The tortoise coordinate $r_*$, defined by $dr_*/dr\equiv r^2/\Delta$, ranges
from $-\infty$ at the horizon to $+\infty$ at spatial infinity.  The
Sasaki-Nakamura equation (\ref{sn}) must be solved imposing an
outgoing-radiation boundary condition at infinity:
\begin{equation}
X(\omega,r)=X^{{\rm out}}e^{\ii \omega r_*} \,, \qquad r_* \rightarrow \infty.
\label{xout}
\end{equation}
The two independent polarization modes of the metric, $h_+$ and $h_{\times}$,
are then given by
\begin{equation}
h_+ + \ii h_{\times}=\frac{8}{r}\int_{-\infty}^{+\infty}d\omega
\sum_{l,m} e^{\ii \omega(r_*-t)} \frac{X^{{\rm
out}}{_{-2}}Y_{lm}(\theta,\phi)}{\lambda(\lambda+2)- 12\ii \omega}\,. 
\label{definitionh}
\end{equation}
Using a formal Green's function solution it is easy to see that $X^{{\rm
out}}\propto {\cal L}$.

Let us now consider the radial infall of an ultrarelativistic point particle
into a non-rotating black hole. The dependence of the emitted gravitational
wave power on the trajectory of the particle comes entirely from the source
${\cal L}$ in eqs.  (\ref{sn})-(\ref{sourceterm}), which in turn depends on
the source term $T$ in Teukolsky's equation (\ref{teukolskyequation}) through
Eq.~(\ref{W}).  To compute the explicit dependence, one would have to consider
generic geodesics in the Kerr geometry \cite{Chandrasekhar:1992bo}, which in
general gives an end result not very amenable to analytical treatments.
However in the highly relativistic regime the equations simplify enormously
\cite{Cardoso:2002ay,Cardoso:2002jr}, and in particular for radial infall
(along the symmetry axis or along the equator) it is possible to find a simple
closed-form analytic expression for the function $W$ in
Eq.~(\ref{sourceterm}). Accordingly, in the following we will provide a
closed-form expressions for ${\cal L}$ in two special situations: radial
infall of an ultrarelativistic particle along the symmetry axis, and radial
infall of an ultrarelativistic particle on the equatorial plane. We will
further specialize to non-rotating black holes, where analytic expressions for
the spin-weighted harmonics are known.

\subsection{Particle falling along the symmetry axis}

Consider a particle falling radially along the symmetry axis of a Kerr black
hole.  In this case the geodesics, written in Boyer-Lindquist coordinates, are
\begin{eqnarray}
\frac{dt}{d\tau}= \epsilon_0\frac{r^2}{\Delta}\,,\quad \, \left
(\frac{dr}{d\tau}\right)^2=\epsilon_0^2-\frac{\Delta}{r^2} \,,\quad
\frac{d\phi}{d\tau}=0\,,
\label{geodesics1}
\end{eqnarray}
where the parameter $\epsilon_0$ is the energy per unit rest mass of the
infalling particle.  On considering highly relativistic particles, $\epsilon_0
\rightarrow \infty$, the source-term function ${\cal L}$ reduces to
\cite{Cardoso:2002ay,Cardoso:2002jr}
\begin{equation}
{\cal L}=-\frac{\mu C \epsilon_0 \gamma_0 \Delta}{2\omega^2
r^5}e^{-\ii \omega r_*}.
\label{explicitS1}
\end{equation}
Here $C=8\left(_{-2}Y_{l0}/\sin^2\theta\right)_{\theta=0}$, the
function $\gamma_0=\gamma_0(r)$ can be found in
\cite{Sasaki:1981sx,Nakamura:1981kk,Sasaki:1981kj,1984PThPh..71...79K,1983PhLA...96..335K,1983PhLA...99...37K,1981PhLA...87...85S},
and $\mu$ is the mass of the particle. We find
\begin{eqnarray}
{\cal L}^{\rm axis}&=&-\frac{\mu \sqrt{\frac{15}{2\pi}}\epsilon_0
\gamma_0 \Delta}{\omega^2 r^5}e^{-\ii \omega
r_*}\,,\quad (l=2)\,,\label{sourceaxisl2}\\
{\cal L}^{\rm axis}&=&-\frac{\mu \sqrt{\frac{105}{2\pi}}\epsilon_0
\gamma_0 \Delta}{\omega^2 r^5}e^{-\ii \omega
r_*}\,,\quad (l=3)\,. \label{explicitS1a0}
\end{eqnarray}

\subsection{Particle falling on the equatorial plane}

Now suppose that the particle is falling radially on the equatorial plane of a
Kerr black hole. On considering highly relativistic particles
we find that the source term takes again a very simple form
\cite{Cardoso:2002ay,Cardoso:2002jr}:
\begin{equation}
{\cal L}=-\frac{\mu \hat{S}\epsilon_0 \gamma_0
\Delta}{\omega^2 r^5}
e^{-\ii \int\frac{K}{\Delta}dr}\,,
\label{explicitS2}
\end{equation}
where
\begin{equation}
\hat{S}=\left[\frac{\lambda}{2}-m^2\right]{_{-2}}Y_{lm}
\left(\pi/2,0\right)
+m\,{_{-2}}Y_{lm}'\left(\pi/2,0\right)\,,
\label{termofontespin}
\end{equation}
and a prime stands for derivative with respect to $\theta$.  For the dominant
multipolar components we get
\begin{subequations}
\begin{eqnarray}
{\cal L}_{20}^{\rm equator}&=&-\frac{\mu \sqrt{\frac{15}{2\pi}}
\epsilon_0 \gamma_0 \Delta}{2\omega^2 r^5}
e^{-\ii \omega r_*}\,,\quad \label{sourceeq20}
{\cal L}_{2\pm2}^{\rm equator}=\frac{3\mu\sqrt{\frac{5}{\pi}}
\epsilon_0 \gamma_0 \Delta}{4\omega^2 r^5}
e^{-\ii \omega r_*}\,,\\
{\cal L}_{3\pm3}^{\rm equator}&=&\pm\frac{5\mu \sqrt{\frac{21}{2\pi}}
\epsilon_0 \gamma_0 \Delta}{4\omega^2 r^5}
e^{-\ii \omega r_*}\,,\quad
{\cal L}_{3\pm1}^{\rm equator}=\mp \frac{3\mu\sqrt{\frac{35}{2\pi}}
\epsilon_0 \gamma_0 \Delta}{4\omega^2 r^5}
e^{-\ii \omega r_*}\,.\label{sourceeq31}
\end{eqnarray}
\end{subequations}
%




\begin{thebibliography}{69}
\expandafter\ifx\csname natexlab\endcsname\relax\def\natexlab#1{#1}\fi
\expandafter\ifx\csname bibnamefont\endcsname\relax
  \def\bibnamefont#1{#1}\fi
\expandafter\ifx\csname bibfnamefont\endcsname\relax
  \def\bibfnamefont#1{#1}\fi
\expandafter\ifx\csname citenamefont\endcsname\relax
  \def\citenamefont#1{#1}\fi
\expandafter\ifx\csname url\endcsname\relax
  \def\url#1{\texttt{#1}}\fi
\expandafter\ifx\csname urlprefix\endcsname\relax\def\urlprefix{URL }\fi
\providecommand{\bibinfo}[2]{#2}
\providecommand{\eprint}[2][]{\url{#2}}

\bibitem[{\citenamefont{Thorne}(1980)}]{Thorne:1980ru}
\bibinfo{author}{\bibfnamefont{K.~S.} \bibnamefont{Thorne}},
  \bibinfo{journal}{Rev. Mod. Phys.} \textbf{\bibinfo{volume}{52}},
  \bibinfo{pages}{299} (\bibinfo{year}{1980}).

\bibitem[{\citenamefont{Newman and Penrose}(1966)}]{Newman:1966ub}
\bibinfo{author}{\bibfnamefont{E.~T.} \bibnamefont{Newman}} \bibnamefont{and}
  \bibinfo{author}{\bibfnamefont{R.}~\bibnamefont{Penrose}},
  \bibinfo{journal}{J. Math. Phys.} \textbf{\bibinfo{volume}{7}},
  \bibinfo{pages}{863} (\bibinfo{year}{1966}).

\bibitem[{\citenamefont{Goldberg et~al.}(1967)\citenamefont{Goldberg,
  MacFarlane, Newman, Rohrlich, and Sudarshan}}]{Goldberg:1966uu}
\bibinfo{author}{\bibfnamefont{J.~N.} \bibnamefont{Goldberg}},
  \bibinfo{author}{\bibfnamefont{A.~J.} \bibnamefont{MacFarlane}},
  \bibinfo{author}{\bibfnamefont{E.~T.} \bibnamefont{Newman}},
  \bibinfo{author}{\bibfnamefont{F.}~\bibnamefont{Rohrlich}}, \bibnamefont{and}
  \bibinfo{author}{\bibfnamefont{E.~C.~G.} \bibnamefont{Sudarshan}},
  \bibinfo{journal}{J. Math. Phys.} \textbf{\bibinfo{volume}{8}},
  \bibinfo{pages}{2155} (\bibinfo{year}{1967}).

\bibitem[{\citenamefont{Baker et~al.}(2007)\citenamefont{Baker, van Meter,
  McWilliams, Centrella, and Kelly}}]{Baker:2006ha}
\bibinfo{author}{\bibfnamefont{J.~G.} \bibnamefont{Baker}},
  \bibinfo{author}{\bibfnamefont{J.~R.} \bibnamefont{van Meter}},
  \bibinfo{author}{\bibfnamefont{S.~T.} \bibnamefont{McWilliams}},
  \bibinfo{author}{\bibfnamefont{J.}~\bibnamefont{Centrella}},
  \bibnamefont{and} \bibinfo{author}{\bibfnamefont{B.~J.} \bibnamefont{Kelly}},
  \bibinfo{journal}{Phys. Rev. Lett.} \textbf{\bibinfo{volume}{99}},
  \bibinfo{pages}{181101} (\bibinfo{year}{2007}), \eprint{gr-qc/0612024}.

\bibitem[{\citenamefont{Pan et~al.}(2008)}]{Pan:2007nw}
\bibinfo{author}{\bibfnamefont{Y.}~\bibnamefont{Pan}} \bibnamefont{et~al.},
  \bibinfo{journal}{Phys. Rev.} \textbf{\bibinfo{volume}{D77}},
  \bibinfo{pages}{024014} (\bibinfo{year}{2008}), \eprint{arXiv:0704.1964
  [gr-qc]}.

\bibitem[{\citenamefont{Buonanno et~al.}(2007{\natexlab{a}})}]{Buonanno:2007pf}
\bibinfo{author}{\bibfnamefont{A.}~\bibnamefont{Buonanno}}
  \bibnamefont{et~al.}, \bibinfo{journal}{Phys. Rev.}
  \textbf{\bibinfo{volume}{D76}}, \bibinfo{pages}{104049}
  (\bibinfo{year}{2007}{\natexlab{a}}), \eprint{arXiv:0706.3732 [gr-qc]}.

\bibitem[{\citenamefont{Boyle et~al.}(2007)}]{Boyle:2007ft}
\bibinfo{author}{\bibfnamefont{M.}~\bibnamefont{Boyle}} \bibnamefont{et~al.},
  \bibinfo{journal}{Phys. Rev.} \textbf{\bibinfo{volume}{D76}},
  \bibinfo{pages}{124038} (\bibinfo{year}{2007}), \eprint{arXiv:0710.0158
  [gr-qc]}.

\bibitem[{\citenamefont{Hannam et~al.}(2008)\citenamefont{Hannam, Husa,
  Sperhake, Brugmann, and Gonzalez}}]{Hannam:2007ik}
\bibinfo{author}{\bibfnamefont{M.}~\bibnamefont{Hannam}},
  \bibinfo{author}{\bibfnamefont{S.}~\bibnamefont{Husa}},
  \bibinfo{author}{\bibfnamefont{U.}~\bibnamefont{Sperhake}},
  \bibinfo{author}{\bibfnamefont{B.}~\bibnamefont{Brugmann}}, \bibnamefont{and}
  \bibinfo{author}{\bibfnamefont{J.~A.} \bibnamefont{Gonzalez}},
  \bibinfo{journal}{Phys. Rev.} \textbf{\bibinfo{volume}{D77}},
  \bibinfo{pages}{044020} (\bibinfo{year}{2008}), \eprint{arXiv:0706.1305
  [gr-qc]}.

\bibitem[{\citenamefont{Damour and Nagar}(2008)}]{Damour:2007yf}
\bibinfo{author}{\bibfnamefont{T.}~\bibnamefont{Damour}} \bibnamefont{and}
  \bibinfo{author}{\bibfnamefont{A.}~\bibnamefont{Nagar}},
  \bibinfo{journal}{Phys. Rev.} \textbf{\bibinfo{volume}{D77}},
  \bibinfo{pages}{024043} (\bibinfo{year}{2008}), \eprint{arXiv:0711.2628
  [gr-qc]}.

\bibitem[{\citenamefont{Ajith et~al.}(2007{\natexlab{a}})}]{Ajith:2007qp}
\bibinfo{author}{\bibfnamefont{P.}~\bibnamefont{Ajith}} \bibnamefont{et~al.},
  \bibinfo{journal}{Class. Quant. Grav.} \textbf{\bibinfo{volume}{24}},
  \bibinfo{pages}{S689} (\bibinfo{year}{2007}{\natexlab{a}}),
  \eprint{arXiv:0704.3764 [gr-qc]}.

\bibitem[{\citenamefont{Ajith et~al.}(2007{\natexlab{b}})}]{Ajith:2007kx}
\bibinfo{author}{\bibfnamefont{P.}~\bibnamefont{Ajith}} \bibnamefont{et~al.}
  (\bibinfo{year}{2007}{\natexlab{b}}), \eprint{arXiv:0710.2335 [gr-qc]}.

\bibitem[{\citenamefont{Yunes and Berti}(2008)}]{Yunes:2008tw}
\bibinfo{author}{\bibfnamefont{N.}~\bibnamefont{Yunes}} \bibnamefont{and}
  \bibinfo{author}{\bibfnamefont{E.}~\bibnamefont{Berti}}
  (\bibinfo{year}{2008}), \eprint{arXiv:0803.1853 [gr-qc]}.

\bibitem[{\citenamefont{Kidder}(2008)}]{Kidder:2007rt}
\bibinfo{author}{\bibfnamefont{L.~E.} \bibnamefont{Kidder}},
  \bibinfo{journal}{Phys. Rev.} \textbf{\bibinfo{volume}{D77}},
  \bibinfo{pages}{044016} (\bibinfo{year}{2008}), \eprint{arXiv:0710.0614
  [gr-qc]}.

\bibitem[{\citenamefont{Blanchet et~al.}(2008)\citenamefont{Blanchet, Faye,
  Iyer, and Sinha}}]{Blanchet:2008je}
\bibinfo{author}{\bibfnamefont{L.}~\bibnamefont{Blanchet}},
  \bibinfo{author}{\bibfnamefont{G.}~\bibnamefont{Faye}},
  \bibinfo{author}{\bibfnamefont{B.~R.} \bibnamefont{Iyer}}, \bibnamefont{and}
  \bibinfo{author}{\bibfnamefont{S.}~\bibnamefont{Sinha}}
  (\bibinfo{year}{2008}), \eprint{arXiv:0802.1249 [gr-qc]}.

\bibitem[{\citenamefont{Buonanno
  et~al.}(2007{\natexlab{b}})\citenamefont{Buonanno, Cook, and
  Pretorius}}]{Buonanno:2006ui}
\bibinfo{author}{\bibfnamefont{A.}~\bibnamefont{Buonanno}},
  \bibinfo{author}{\bibfnamefont{G.~B.} \bibnamefont{Cook}}, \bibnamefont{and}
  \bibinfo{author}{\bibfnamefont{F.}~\bibnamefont{Pretorius}},
  \bibinfo{journal}{Phys. Rev.} \textbf{\bibinfo{volume}{D75}},
  \bibinfo{pages}{124018} (\bibinfo{year}{2007}{\natexlab{b}}),
  \eprint{gr-qc/0610122}.

\bibitem[{\citenamefont{Berti et~al.}(2007{\natexlab{a}})}]{Berti:2007fi}
\bibinfo{author}{\bibfnamefont{E.}~\bibnamefont{Berti}} \bibnamefont{et~al.},
  \bibinfo{journal}{Phys. Rev.} \textbf{\bibinfo{volume}{D76}},
  \bibinfo{pages}{064034} (\bibinfo{year}{2007}{\natexlab{a}}),
  \eprint{gr-qc/0703053}.

\bibitem[{\citenamefont{Schnittman et~al.}(2008)}]{Schnittman:2007ij}
\bibinfo{author}{\bibfnamefont{J.~D.} \bibnamefont{Schnittman}}
  \bibnamefont{et~al.}, \bibinfo{journal}{Phys. Rev.}
  \textbf{\bibinfo{volume}{D77}}, \bibinfo{pages}{044031}
  (\bibinfo{year}{2008}), \eprint{arXiv:0707.0301 [gr-qc]}.

\bibitem[{\citenamefont{Vaishnav et~al.}(2007)\citenamefont{Vaishnav, Hinder,
  Herrmann, and Shoemaker}}]{Vaishnav:2007nm}
\bibinfo{author}{\bibfnamefont{B.}~\bibnamefont{Vaishnav}},
  \bibinfo{author}{\bibfnamefont{I.}~\bibnamefont{Hinder}},
  \bibinfo{author}{\bibfnamefont{F.}~\bibnamefont{Herrmann}}, \bibnamefont{and}
  \bibinfo{author}{\bibfnamefont{D.}~\bibnamefont{Shoemaker}},
  \bibinfo{journal}{Phys. Rev.} \textbf{\bibinfo{volume}{D76}},
  \bibinfo{pages}{084020} (\bibinfo{year}{2007}), \eprint{arXiv:0705.3829
  [gr-qc]}.

\bibitem[{\citenamefont{Berti et~al.}(2007{\natexlab{b}})\citenamefont{Berti,
  Cardoso, Gonzalez, Sperhake, and Brugmann}}]{Berti:2007nw}
\bibinfo{author}{\bibfnamefont{E.}~\bibnamefont{Berti}},
  \bibinfo{author}{\bibfnamefont{V.}~\bibnamefont{Cardoso}},
  \bibinfo{author}{\bibfnamefont{J.~A.} \bibnamefont{Gonzalez}},
  \bibinfo{author}{\bibfnamefont{U.}~\bibnamefont{Sperhake}}, \bibnamefont{and}
  \bibinfo{author}{\bibfnamefont{B.}~\bibnamefont{Brugmann}}
  (\bibinfo{year}{2007}{\natexlab{b}}), \eprint{arXiv:0711.1097 [gr-qc]}.

\bibitem[{\citenamefont{Bondi et~al.}(1962)\citenamefont{Bondi, van~der Burg,
  and Metzner}}]{Bondi:1962px}
\bibinfo{author}{\bibfnamefont{H.}~\bibnamefont{Bondi}},
  \bibinfo{author}{\bibfnamefont{M.~G.~J.} \bibnamefont{van~der Burg}},
  \bibnamefont{and} \bibinfo{author}{\bibfnamefont{A.~W.~K.}
  \bibnamefont{Metzner}}, \bibinfo{journal}{Proc. Roy. Soc. Lond.}
  \textbf{\bibinfo{volume}{A269}}, \bibinfo{pages}{21} (\bibinfo{year}{1962}).

\bibitem[{\citenamefont{Sachs}(1962{\natexlab{a}})}]{Sachs:1962zza}
\bibinfo{author}{\bibfnamefont{R.}~\bibnamefont{Sachs}},
  \bibinfo{journal}{Phys. Rev.} \textbf{\bibinfo{volume}{128}},
  \bibinfo{pages}{2851} (\bibinfo{year}{1962}{\natexlab{a}}).

\bibitem[{\citenamefont{Sachs}(1962{\natexlab{b}})}]{Sachs:1962wk}
\bibinfo{author}{\bibfnamefont{R.~K.} \bibnamefont{Sachs}},
  \bibinfo{journal}{Proc. Roy. Soc. Lond.} \textbf{\bibinfo{volume}{A270}},
  \bibinfo{pages}{103} (\bibinfo{year}{1962}{\natexlab{b}}).

\bibitem[{\citenamefont{Campanelli
  et~al.}(2007{\natexlab{a}})\citenamefont{Campanelli, Lousto, Zlochower, and
  Merritt}}]{Campanelli:2007ew}
\bibinfo{author}{\bibfnamefont{M.}~\bibnamefont{Campanelli}},
  \bibinfo{author}{\bibfnamefont{C.~O.} \bibnamefont{Lousto}},
  \bibinfo{author}{\bibfnamefont{Y.}~\bibnamefont{Zlochower}},
  \bibnamefont{and} \bibinfo{author}{\bibfnamefont{D.}~\bibnamefont{Merritt}},
  \bibinfo{journal}{Astrophys. J.} \textbf{\bibinfo{volume}{659}},
  \bibinfo{pages}{L5} (\bibinfo{year}{2007}{\natexlab{a}}),
  \eprint{gr-qc/0701164}.

\bibitem[{\citenamefont{Campanelli
  et~al.}(2007{\natexlab{b}})\citenamefont{Campanelli, Lousto, Zlochower, and
  Merritt}}]{Campanelli:2007cga}
\bibinfo{author}{\bibfnamefont{M.}~\bibnamefont{Campanelli}},
  \bibinfo{author}{\bibfnamefont{C.~O.} \bibnamefont{Lousto}},
  \bibinfo{author}{\bibfnamefont{Y.}~\bibnamefont{Zlochower}},
  \bibnamefont{and} \bibinfo{author}{\bibfnamefont{D.}~\bibnamefont{Merritt}},
  \bibinfo{journal}{Phys. Rev. Lett.} \textbf{\bibinfo{volume}{98}},
  \bibinfo{pages}{231102} (\bibinfo{year}{2007}{\natexlab{b}}),
  \eprint{gr-qc/0702133}.

\bibitem[{\citenamefont{Gonzalez et~al.}(2007)\citenamefont{Gonzalez, Hannam,
  Sperhake, Brugmann, and Husa}}]{Gonzalez:2007hi}
\bibinfo{author}{\bibfnamefont{J.~A.} \bibnamefont{Gonzalez}},
  \bibinfo{author}{\bibfnamefont{M.~D.} \bibnamefont{Hannam}},
  \bibinfo{author}{\bibfnamefont{U.}~\bibnamefont{Sperhake}},
  \bibinfo{author}{\bibfnamefont{B.}~\bibnamefont{Brugmann}}, \bibnamefont{and}
  \bibinfo{author}{\bibfnamefont{S.}~\bibnamefont{Husa}},
  \bibinfo{journal}{Phys. Rev. Lett.} \textbf{\bibinfo{volume}{98}},
  \bibinfo{pages}{231101} (\bibinfo{year}{2007}), \eprint{gr-qc/0702052}.

\bibitem[{\citenamefont{Pollney et~al.}(2007)}]{Pollney:2007ss}
\bibinfo{author}{\bibfnamefont{D.}~\bibnamefont{Pollney}} \bibnamefont{et~al.},
  \bibinfo{journal}{Phys. Rev.} \textbf{\bibinfo{volume}{D76}},
  \bibinfo{pages}{124002} (\bibinfo{year}{2007}).

\bibitem[{\citenamefont{Dain et~al.}(2008)\citenamefont{Dain, Lousto, and
  Zlochower}}]{Dain:2008ck}
\bibinfo{author}{\bibfnamefont{S.}~\bibnamefont{Dain}},
  \bibinfo{author}{\bibfnamefont{C.~O.} \bibnamefont{Lousto}},
  \bibnamefont{and} \bibinfo{author}{\bibfnamefont{Y.}~\bibnamefont{Zlochower}}
  (\bibinfo{year}{2008}), \eprint{arXiv:0803.0351 [gr-qc]}.

\bibitem[{\citenamefont{Flanagan and Hughes}(1998)}]{Flanagan:1997sx}
\bibinfo{author}{\bibfnamefont{E.~E.} \bibnamefont{Flanagan}} \bibnamefont{and}
  \bibinfo{author}{\bibfnamefont{S.~A.} \bibnamefont{Hughes}},
  \bibinfo{journal}{Phys. Rev.} \textbf{\bibinfo{volume}{D57}},
  \bibinfo{pages}{4535} (\bibinfo{year}{1998}), \eprint{gr-qc/9701039}.

\bibitem[{\citenamefont{Berti et~al.}(2007{\natexlab{c}})\citenamefont{Berti,
  Cardoso, Cardoso, and Cavaglia}}]{Berti:2007zu}
\bibinfo{author}{\bibfnamefont{E.}~\bibnamefont{Berti}},
  \bibinfo{author}{\bibfnamefont{J.}~\bibnamefont{Cardoso}},
  \bibinfo{author}{\bibfnamefont{V.}~\bibnamefont{Cardoso}}, \bibnamefont{and}
  \bibinfo{author}{\bibfnamefont{M.}~\bibnamefont{Cavaglia}},
  \bibinfo{journal}{Phys. Rev.} \textbf{\bibinfo{volume}{D76}},
  \bibinfo{pages}{104044} (\bibinfo{year}{2007}{\natexlab{c}}),
  \eprint{arXiv:0707.1202 [gr-qc]}.

\bibitem[{\citenamefont{Brown et~al.}(2007)}]{Brown:2007jx}
\bibinfo{author}{\bibfnamefont{D.}~\bibnamefont{Brown}} \bibnamefont{et~al.}
  (\bibinfo{year}{2007}), \eprint{arXiv:0709.0093 [gr-qc]}.

\bibitem[{nin()}]{ninja}
\bibinfo{howpublished}{\url{http://www.gravity.phy.syr.edu/dokuwiki/doku.php?i%
d=ninja:home}}.

\bibitem[{\citenamefont{Berti et~al.}(2006)\citenamefont{Berti, Cardoso, and
  Will}}]{Berti:2005ys}
\bibinfo{author}{\bibfnamefont{E.}~\bibnamefont{Berti}},
  \bibinfo{author}{\bibfnamefont{V.}~\bibnamefont{Cardoso}}, \bibnamefont{and}
  \bibinfo{author}{\bibfnamefont{C.~M.} \bibnamefont{Will}},
  \bibinfo{journal}{Phys. Rev.} \textbf{\bibinfo{volume}{D73}},
  \bibinfo{pages}{064030} (\bibinfo{year}{2006}), \eprint{gr-qc/0512160}.

\bibitem[{\citenamefont{Sperhake et~al.}(2008)\citenamefont{Sperhake, Cardoso,
  Pretorius, Berti, and Gonzalez}}]{Sperhake:2008ga}
\bibinfo{author}{\bibfnamefont{U.}~\bibnamefont{Sperhake}},
  \bibinfo{author}{\bibfnamefont{V.}~\bibnamefont{Cardoso}},
  \bibinfo{author}{\bibfnamefont{F.}~\bibnamefont{Pretorius}},
  \bibinfo{author}{\bibfnamefont{E.}~\bibnamefont{Berti}}, \bibnamefont{and}
  \bibinfo{author}{\bibfnamefont{J.~A.} \bibnamefont{Gonzalez}}
  (\bibinfo{year}{2008}), \eprint{arXiv:0806.1738 [gr-qc]}.

\bibitem[{\citenamefont{McCarthy}(1972)}]{mccarthy:1837}
\bibinfo{author}{\bibfnamefont{P.~J.} \bibnamefont{McCarthy}},
  \bibinfo{journal}{Journal of Mathematical Physics}
  \textbf{\bibinfo{volume}{13}}, \bibinfo{pages}{1837} (\bibinfo{year}{1972}).

\bibitem[{\citenamefont{Wald}(1984)}]{Wald:1984cw}
\bibinfo{author}{\bibfnamefont{R.}~\bibnamefont{Wald}},
  \emph{\bibinfo{title}{General Relativity}} (\bibinfo{publisher}{The
  University of Chicago press}, \bibinfo{address}{Chicago},
  \bibinfo{year}{1984}).

\bibitem[{\citenamefont{Carmeli}(2000)}]{Carmeli:2000af}
\bibinfo{author}{\bibfnamefont{M.}~\bibnamefont{Carmeli}},
  \emph{\bibinfo{title}{{Group theory and general relativity: Representations
  of the Lorentz group and their applications to the gravitational field}}}
  (\bibinfo{publisher}{World Scientific}, \bibinfo{address}{Singapore},
  \bibinfo{year}{2000}).

\bibitem[{\citenamefont{Newman and Unti}(1962)}]{newman:891}
\bibinfo{author}{\bibfnamefont{E.~T.} \bibnamefont{Newman}} \bibnamefont{and}
  \bibinfo{author}{\bibfnamefont{T.~W.~J.} \bibnamefont{Unti}},
  \bibinfo{journal}{Journal of Mathematical Physics}
  \textbf{\bibinfo{volume}{3}}, \bibinfo{pages}{891} (\bibinfo{year}{1962}).

\bibitem[{\citenamefont{Geroch}(1977)}]{Geroch1977}
\bibinfo{author}{\bibfnamefont{R.}~\bibnamefont{Geroch}}, in
  \emph{\bibinfo{booktitle}{Asymptotic Structure of Space-Time}}, edited by
  \bibinfo{editor}{\bibfnamefont{F.~P.} \bibnamefont{Esposito}}
  \bibnamefont{and} \bibinfo{editor}{\bibfnamefont{L.}~\bibnamefont{Witten}}
  (\bibinfo{publisher}{Plenum Press}, \bibinfo{address}{New York},
  \bibinfo{year}{1977}).

\bibitem[{\citenamefont{Newman and Penrose}(1962)}]{newman:566}
\bibinfo{author}{\bibfnamefont{E.}~\bibnamefont{Newman}} \bibnamefont{and}
  \bibinfo{author}{\bibfnamefont{R.}~\bibnamefont{Penrose}},
  \bibinfo{journal}{Journal of Mathematical Physics}
  \textbf{\bibinfo{volume}{3}}, \bibinfo{pages}{566} (\bibinfo{year}{1962}).

\bibitem[{\citenamefont{Lehner and Moreschi}(2007)}]{Lehner:2007ip}
\bibinfo{author}{\bibfnamefont{L.}~\bibnamefont{Lehner}} \bibnamefont{and}
  \bibinfo{author}{\bibfnamefont{O.~M.} \bibnamefont{Moreschi}},
  \bibinfo{journal}{Phys. Rev.} \textbf{\bibinfo{volume}{D76}},
  \bibinfo{pages}{124040} (\bibinfo{year}{2007}), \eprint{arXiv:0706.1319
  [gr-qc]}.

\bibitem[{\citenamefont{{Teukolsky}}(1973)}]{Teukolsky:1973ap}
\bibinfo{author}{\bibfnamefont{S.~A.} \bibnamefont{{Teukolsky}}},
  \bibinfo{journal}{\apj} \textbf{\bibinfo{volume}{185}}, \bibinfo{pages}{635}
  (\bibinfo{year}{1973}).

\bibitem[{\citenamefont{Brugmann et~al.}(2008)}]{Brugmann:2008zz}
\bibinfo{author}{\bibfnamefont{B.}~\bibnamefont{Brugmann}}
  \bibnamefont{et~al.}, \bibinfo{journal}{Phys. Rev.}
  \textbf{\bibinfo{volume}{D77}}, \bibinfo{pages}{024027}
  (\bibinfo{year}{2008}), \eprint{gr-qc/0610128}.

\bibitem[{\citenamefont{Sperhake}(2007)}]{Sperhake:2006cy}
\bibinfo{author}{\bibfnamefont{U.}~\bibnamefont{Sperhake}},
  \bibinfo{journal}{Phys. Rev.} \textbf{\bibinfo{volume}{D76}},
  \bibinfo{pages}{104015} (\bibinfo{year}{2007}), \eprint{gr-qc/0606079}.

\bibitem[{\citenamefont{Misner et~al.}(1973)\citenamefont{Misner, Thorne, and
  Wheeler}}]{Misner:1973cw}
\bibinfo{author}{\bibfnamefont{C.~W.} \bibnamefont{Misner}},
  \bibinfo{author}{\bibfnamefont{K.}~\bibnamefont{Thorne}}, \bibnamefont{and}
  \bibinfo{author}{\bibfnamefont{J.~A.} \bibnamefont{Wheeler}},
  \emph{\bibinfo{title}{Gravitation}} (\bibinfo{publisher}{W. H. Freeman \&
  Co.}, \bibinfo{address}{San Francisco}, \bibinfo{year}{1973}).

\bibitem[{\citenamefont{Brown et~al.}(1997)\citenamefont{Brown, Lau, and
  York}}]{Brown:1996bw}
\bibinfo{author}{\bibfnamefont{J.~D.} \bibnamefont{Brown}},
  \bibinfo{author}{\bibfnamefont{S.~R.} \bibnamefont{Lau}}, \bibnamefont{and}
  \bibinfo{author}{\bibfnamefont{J.}~\bibnamefont{York},
  \bibfnamefont{James~W.}}, \bibinfo{journal}{Phys. Rev.}
  \textbf{\bibinfo{volume}{D55}}, \bibinfo{pages}{1977} (\bibinfo{year}{1997}),
  \eprint{gr-qc/9609057}.

\bibitem[{\citenamefont{Sperhake et~al.}(2007)}]{Sperhake:2007gu}
\bibinfo{author}{\bibfnamefont{U.}~\bibnamefont{Sperhake}} \bibnamefont{et~al.}
  (\bibinfo{year}{2007}), \eprint{arXiv:0710.3823 [gr-qc]}.

\bibitem[{\citenamefont{Palenzuela et~al.}(2007)\citenamefont{Palenzuela,
  Olabarrieta, Lehner, and Liebling}}]{Palenzuela:2006wp}
\bibinfo{author}{\bibfnamefont{C.}~\bibnamefont{Palenzuela}},
  \bibinfo{author}{\bibfnamefont{I.}~\bibnamefont{Olabarrieta}},
  \bibinfo{author}{\bibfnamefont{L.}~\bibnamefont{Lehner}}, \bibnamefont{and}
  \bibinfo{author}{\bibfnamefont{S.}~\bibnamefont{Liebling}},
  \bibinfo{journal}{Phys. Rev.} \textbf{\bibinfo{volume}{D75}},
  \bibinfo{pages}{064005} (\bibinfo{year}{2007}), \eprint{gr-qc/0612067}.

\bibitem[{\citenamefont{Eardley
  et~al.}(1973{\natexlab{a}})\citenamefont{Eardley, Lee, Lightman, Wagoner, and
  Will}}]{Eardley:1973br}
\bibinfo{author}{\bibfnamefont{D.~M.} \bibnamefont{Eardley}},
  \bibinfo{author}{\bibfnamefont{D.~L.} \bibnamefont{Lee}},
  \bibinfo{author}{\bibfnamefont{A.~P.} \bibnamefont{Lightman}},
  \bibinfo{author}{\bibfnamefont{R.~V.} \bibnamefont{Wagoner}},
  \bibnamefont{and} \bibinfo{author}{\bibfnamefont{C.~M.} \bibnamefont{Will}},
  \bibinfo{journal}{Phys. Rev. Lett.} \textbf{\bibinfo{volume}{30}},
  \bibinfo{pages}{884} (\bibinfo{year}{1973}{\natexlab{a}}).

\bibitem[{\citenamefont{Eardley
  et~al.}(1973{\natexlab{b}})\citenamefont{Eardley, Lee, and
  Lightman}}]{Eardley:1974nw}
\bibinfo{author}{\bibfnamefont{D.~M.} \bibnamefont{Eardley}},
  \bibinfo{author}{\bibfnamefont{D.~L.} \bibnamefont{Lee}}, \bibnamefont{and}
  \bibinfo{author}{\bibfnamefont{A.~P.} \bibnamefont{Lightman}},
  \bibinfo{journal}{Phys. Rev.} \textbf{\bibinfo{volume}{D8}},
  \bibinfo{pages}{3308} (\bibinfo{year}{1973}{\natexlab{b}}).

\bibitem[{\citenamefont{Alexander et~al.}(2007)\citenamefont{Alexander, Finn,
  and Yunes}}]{Alexander:2007kv}
\bibinfo{author}{\bibfnamefont{S.}~\bibnamefont{Alexander}},
  \bibinfo{author}{\bibfnamefont{L.~S.} \bibnamefont{Finn}}, \bibnamefont{and}
  \bibinfo{author}{\bibfnamefont{N.}~\bibnamefont{Yunes}}
  (\bibinfo{year}{2007}), \eprint{arXiv:0712.2542 [gr-qc]}.

\bibitem[{\citenamefont{Berti et~al.}(2007{\natexlab{d}})\citenamefont{Berti,
  Cardoso, Gonzalez, and Sperhake}}]{Berti:2007dg}
\bibinfo{author}{\bibfnamefont{E.}~\bibnamefont{Berti}},
  \bibinfo{author}{\bibfnamefont{V.}~\bibnamefont{Cardoso}},
  \bibinfo{author}{\bibfnamefont{J.~A.} \bibnamefont{Gonzalez}},
  \bibnamefont{and} \bibinfo{author}{\bibfnamefont{U.}~\bibnamefont{Sperhake}},
  \bibinfo{journal}{Phys. Rev.} \textbf{\bibinfo{volume}{D75}},
  \bibinfo{pages}{124017} (\bibinfo{year}{2007}{\natexlab{d}}),
  \eprint{arXiv:0701086 [gr-qc]}.

\bibitem[{\citenamefont{Campanelli and Lousto}(1999)}]{Campanelli:1998jv}
\bibinfo{author}{\bibfnamefont{M.}~\bibnamefont{Campanelli}} \bibnamefont{and}
  \bibinfo{author}{\bibfnamefont{C.~O.} \bibnamefont{Lousto}},
  \bibinfo{journal}{Phys. Rev.} \textbf{\bibinfo{volume}{D59}},
  \bibinfo{pages}{124022} (\bibinfo{year}{1999}), \eprint{gr-qc/9811019}.

\bibitem[{\citenamefont{Lousto and Zlochower}(2007)}]{Lousto:2007mh}
\bibinfo{author}{\bibfnamefont{C.~O.} \bibnamefont{Lousto}} \bibnamefont{and}
  \bibinfo{author}{\bibfnamefont{Y.}~\bibnamefont{Zlochower}},
  \bibinfo{journal}{Phys. Rev.} \textbf{\bibinfo{volume}{D76}},
  \bibinfo{pages}{041502} (\bibinfo{year}{2007}), \eprint{gr-qc/0703061}.

\bibitem[{\citenamefont{Ruiz et~al.}(2007)\citenamefont{Ruiz, Takahashi,
  Alcubierre, and Nunez}}]{Ruiz:2007yx}
\bibinfo{author}{\bibfnamefont{M.}~\bibnamefont{Ruiz}},
  \bibinfo{author}{\bibfnamefont{R.}~\bibnamefont{Takahashi}},
  \bibinfo{author}{\bibfnamefont{M.}~\bibnamefont{Alcubierre}},
  \bibnamefont{and} \bibinfo{author}{\bibfnamefont{D.}~\bibnamefont{Nunez}}
  (\bibinfo{year}{2007}), \eprint{arXiv:0707.4654 [gr-qc]}.

\bibitem[{\citenamefont{Baker et~al.}(2006)\citenamefont{Baker, Centrella,
  Choi, Koppitz, and van Meter}}]{Baker:2006yw}
\bibinfo{author}{\bibfnamefont{J.~G.} \bibnamefont{Baker}},
  \bibinfo{author}{\bibfnamefont{J.}~\bibnamefont{Centrella}},
  \bibinfo{author}{\bibfnamefont{D.-I.} \bibnamefont{Choi}},
  \bibinfo{author}{\bibfnamefont{M.}~\bibnamefont{Koppitz}}, \bibnamefont{and}
  \bibinfo{author}{\bibfnamefont{J.}~\bibnamefont{van Meter}},
  \bibinfo{journal}{Phys. Rev.} \textbf{\bibinfo{volume}{D73}},
  \bibinfo{pages}{104002} (\bibinfo{year}{2006}), \eprint{gr-qc/0602026}.

\bibitem[{\citenamefont{{Teukolsky}}(1972)}]{Teukolsky:1972le}
\bibinfo{author}{\bibfnamefont{S.~A.} \bibnamefont{{Teukolsky}}},
  \bibinfo{journal}{\prl} \textbf{\bibinfo{volume}{29}}, \bibinfo{pages}{1114}
  (\bibinfo{year}{1972}).

\bibitem[{\citenamefont{Sasaki and
  Nakamura}(1982{\natexlab{a}})}]{Sasaki:1981sx}
\bibinfo{author}{\bibfnamefont{M.}~\bibnamefont{Sasaki}} \bibnamefont{and}
  \bibinfo{author}{\bibfnamefont{T.}~\bibnamefont{Nakamura}},
  \bibinfo{journal}{Prog. Theor. Phys.} \textbf{\bibinfo{volume}{67}},
  \bibinfo{pages}{1788} (\bibinfo{year}{1982}{\natexlab{a}}).

\bibitem[{\citenamefont{Nakamura and Sasaki}(1982)}]{Nakamura:1981kk}
\bibinfo{author}{\bibfnamefont{T.}~\bibnamefont{Nakamura}} \bibnamefont{and}
  \bibinfo{author}{\bibfnamefont{M.}~\bibnamefont{Sasaki}},
  \bibinfo{journal}{Phys. Lett.} \textbf{\bibinfo{volume}{A89}},
  \bibinfo{pages}{185} (\bibinfo{year}{1982}).

\bibitem[{\citenamefont{Sasaki and
  Nakamura}(1982{\natexlab{b}})}]{Sasaki:1981kj}
\bibinfo{author}{\bibfnamefont{M.}~\bibnamefont{Sasaki}} \bibnamefont{and}
  \bibinfo{author}{\bibfnamefont{T.}~\bibnamefont{Nakamura}},
  \bibinfo{journal}{Phys. Lett.} \textbf{\bibinfo{volume}{A89}},
  \bibinfo{pages}{68} (\bibinfo{year}{1982}{\natexlab{b}}).

\bibitem[{\citenamefont{{Kojima} and {Nakamura}}(1984)}]{1984PThPh..71...79K}
\bibinfo{author}{\bibfnamefont{Y.}~\bibnamefont{{Kojima}}} \bibnamefont{and}
  \bibinfo{author}{\bibfnamefont{T.}~\bibnamefont{{Nakamura}}},
  \bibinfo{journal}{Progress of Theoretical Physics}
  \textbf{\bibinfo{volume}{71}}, \bibinfo{pages}{79} (\bibinfo{year}{1984}).

\bibitem[{\citenamefont{{Kojima} and
  {Nakamura}}(1983{\natexlab{a}})}]{1983PhLA...96..335K}
\bibinfo{author}{\bibfnamefont{Y.}~\bibnamefont{{Kojima}}} \bibnamefont{and}
  \bibinfo{author}{\bibfnamefont{T.}~\bibnamefont{{Nakamura}}},
  \bibinfo{journal}{Physics Letters A} \textbf{\bibinfo{volume}{96}},
  \bibinfo{pages}{335} (\bibinfo{year}{1983}{\natexlab{a}}).

\bibitem[{\citenamefont{{Kojima} and
  {Nakamura}}(1983{\natexlab{b}})}]{1983PhLA...99...37K}
\bibinfo{author}{\bibfnamefont{Y.}~\bibnamefont{{Kojima}}} \bibnamefont{and}
  \bibinfo{author}{\bibfnamefont{T.}~\bibnamefont{{Nakamura}}},
  \bibinfo{journal}{Physics Letters A} \textbf{\bibinfo{volume}{99}},
  \bibinfo{pages}{37} (\bibinfo{year}{1983}{\natexlab{b}}).

\bibitem[{\citenamefont{{Sasaki} and {Nakamura}}(1981)}]{1981PhLA...87...85S}
\bibinfo{author}{\bibfnamefont{M.}~\bibnamefont{{Sasaki}}} \bibnamefont{and}
  \bibinfo{author}{\bibfnamefont{T.}~\bibnamefont{{Nakamura}}},
  \bibinfo{journal}{Physics Letters A} \textbf{\bibinfo{volume}{87}},
  \bibinfo{pages}{85} (\bibinfo{year}{1981}).

\bibitem[{\citenamefont{Hughes}(2000)}]{Hughes:1999bq}
\bibinfo{author}{\bibfnamefont{S.~A.} \bibnamefont{Hughes}},
  \bibinfo{journal}{Phys. Rev.} \textbf{\bibinfo{volume}{D61}},
  \bibinfo{pages}{084004} (\bibinfo{year}{2000}), \eprint{gr-qc/9910091}.

\bibitem[{\citenamefont{Kennefick}(1998)}]{Kennefick:1998ab}
\bibinfo{author}{\bibfnamefont{D.}~\bibnamefont{Kennefick}},
  \bibinfo{journal}{Phys. Rev.} \textbf{\bibinfo{volume}{D58}},
  \bibinfo{pages}{064012} (\bibinfo{year}{1998}), \eprint{gr-qc/9805102}.

\bibitem[{\citenamefont{Cardoso and Lemos}(2002)}]{Cardoso:2002ay}
\bibinfo{author}{\bibfnamefont{V.}~\bibnamefont{Cardoso}} \bibnamefont{and}
  \bibinfo{author}{\bibfnamefont{J.~P.~S.} \bibnamefont{Lemos}},
  \bibinfo{journal}{Phys. Lett.} \textbf{\bibinfo{volume}{B538}},
  \bibinfo{pages}{1} (\bibinfo{year}{2002}), \eprint{gr-qc/0202019}.

\bibitem[{\citenamefont{Cardoso and Lemos}(2003)}]{Cardoso:2002jr}
\bibinfo{author}{\bibfnamefont{V.}~\bibnamefont{Cardoso}} \bibnamefont{and}
  \bibinfo{author}{\bibfnamefont{J.~P.~S.} \bibnamefont{Lemos}},
  \bibinfo{journal}{Phys. Rev.} \textbf{\bibinfo{volume}{D67}},
  \bibinfo{pages}{084005} (\bibinfo{year}{2003}), \eprint{gr-qc/0211094}.

\bibitem[{\citenamefont{{Breuer}}(1975)}]{1975LNP....44.....B}
\bibinfo{editor}{\bibfnamefont{R.~A.} \bibnamefont{{Breuer}}}, ed.,
  \emph{\bibinfo{title}{{Gravitational perturbation theory and synchrotron
  radiation}}}, vol.~\bibinfo{volume}{44} of \emph{\bibinfo{series}{Lecture
  Notes in Physics, Berlin Springer Verlag}} (\bibinfo{year}{1975}).

\bibitem[{\citenamefont{{Chandrasekhar}}(1992)}]{Chandrasekhar:1992bo}
\bibinfo{author}{\bibfnamefont{S.}~\bibnamefont{{Chandrasekhar}}},
  \emph{\bibinfo{title}{{The mathematical theory of black holes}}}
  (\bibinfo{publisher}{Oxford University Press}, \bibinfo{address}{New York},
  \bibinfo{year}{1992}).

\end{thebibliography}
\end{document}